\pdfoutput=1
\newcommand{\FullVersion}{}
\ifdefined\FullVersion
    \documentclass[acmsmall,nonacm]{acmart}
\else
    \documentclass[acmsmall]{acmart}
\fi
\usepackage{00}

\acmJournal{PACMPL}
\acmVolume{1}
\acmNumber{CONF}
\acmArticle{1}
\acmYear{2018}
\acmMonth{1}
\acmDOI{}
\startPage{1}

\setcopyright{none}

\begin{document}

\ifdefined\FullVersion
    \title{Fluent APIs in Functional Languages (full version)}
\else
    \title{Fluent APIs in Functional Languages}
\fi

\ifdefined\FullVersion\else
\titlenote{See a full version of this article \cite{Roth:2023} and the accompanying artifact \cite{Roth:2023:artifact}.}
\fi

\author{Ori Roth}
\email{soriroth@cs.technion.ac.il}
\author{Yossi Gil}
\email{yogi@cs.technion.ac.il}
\affiliation{%
	\institution{Technion---IIT}
	\city{Haifa}
	\country{Israel}}

\begin{abstract}
Fluent API is an object-oriented pattern for elegant APIs and embedded DSLs.
A smart fluent API can enforce the API protocol or DSL syntax at compile time.
As fluent API designs typically rely on function overloading, they are hard to realize in functional programming languages.
We show how to write functional fluent APIs using parametric polymorphism and type inference instead of overloading.
Our designs support all regular and deterministic context-free API protocols and beyond.

\end{abstract}



\maketitle

\section{Introduction}
\label{section:aa}
An API protocol dictates a set of rules for correct API usage.
Misusing an API to contrast its protocol may result in runtime errors or undefined behavior.
The Java \inline{FileReader} API protocol, for example, states that
\begin{quote}
	``once the stream has been closed, further [API] invocations will throw an \inline{IOException}.''\footnote{\url{https://docs.oracle.com/javase/8/docs/api/java/io/InputStreamReader.html\#close--}}
\end{quote}
Although protocols usually come in written form as part of the API documentation, some protocols can be formalized as grammars.
Such grammar defines the language (set) of legal sequences of API invocations.
For example, the following regular expression describes the \inline{FileReader} protocol:
\begin{quote}
	\centering
	$\text{\inline{$\langle$initialize$\rangle$}}\;\;\;\big(\;\text{\inline{read}}~\big|~\text{\inline{skip}}~\big|~\text{\inline{reset}}\;\big)^*\;\;\;\text{\inline{close}}^*$
\end{quote}
Specifying API protocols using grammars or automata reduces the problem of protocol verification to formal language recognition.

APIs also use grammars and automata to specify domain-specific languages (DSLs).
A DSL is a specialized language tailored for the API's problem domain.
Many applications use DSLs:
A Python machine learning model uses SQL to load its dataset, a JavaScript frontend verifies textual forms with regular expressions, and a Standard ML (SML) HTTP server serves HTML webpages.
To integrate a DSL into an application, we can implement the DSL as an API in the host programming language.
The protocol of this embedded DSL (EDSL) is the syntax of the original DSL, i.e., EDSL invocations must describe well-formed DSL programs.

This paper presents methods to create \emph{smart} and \emph{elegant} APIs in functional programming languages.
These APIs enforce their protocols or DSL syntax at compile time by performing language recognition at the type level.
For example, \cref{listing:html} demonstrates how we embed an HTML document as an SML expression using our methodology.

\begin{JAVA}[float=ht,style=sml,caption={An HTML webpage embedded in SML},label={listing:html}]
val webpage = ^^
    <html>
    	<body>
    		<h1> `"National Parks" </h1>
    		`"California:"
    		<table>
    			<tr>
    				<th> `"Park Description" </th>
    				<th> `"Park Picture" </th>
    			</tr>
    			<tr>
    				<td> <p> <b> `"Yosemite" </b> `"national park" </p> </td>
    				<td> <img src "https://tinyurl.com/yosemite5"/> </td>
    			</tr>
    		</table>
    	</body>
    </html>
\$\$
\end{JAVA}

Our HTML EDSL is \emph{elegant} in the sense that embedded webpages closely resemble HTML, with minimal syntactic overhead imposed by the host programming language.
The EDSL is \emph{smart} as it coerces the SML compiler into enforcing HTML's syntax, so invalid webpages do not compile.
The tags must appear in the correct order (\inline{<html>} before \inline{<body>}, \inline{<tr>} inside \inline{<table>}, and so on), every opening tag must have a matching closing tag, and all table rows must have the same number of columns.
Note that the additional requirement on table rows elevates HTML into a context-sensitive language.

Our API designs are based on \emph{fluent API}, an object-oriented pattern for embedding DSLs.
Fluent API techniques typically rely on function overloading, a common object-oriented feature not supported by many functional type systems.
We adapt fluent APIs to functional settings by replacing overloading with parametric polymorphism and type inference.
Our fluent APIs do not rely on a specific language feature, so they are compatible with most statically-typed functional languages.

\subsection{Related Work}

A typestate \cite{Strom:1986} determines the type's set of permitted operations in a given program context.
The \inline{FileReader} typestate, for example, disallows invoking \inline{read} after \inline{close} is called.
Typestates are used as a first-class abstraction for protocols \cite{Aldrich:2009,Garcia:2014,Sunshine:2011,DeLine:2001,Degen:2007,Kuncak:2002}.
Another approach is to infer and verify typestates and API protocols statically \cite{Fink:2008,Bodden:2012,Field:2003,Pradel:2012}.
\citet{Ferles:2021} developed a static analysis method for enforcing API protocols specified by context-free grammars.
Similarly, the current work uses automata specifications for API protocols.
Instead of using an external tool or a dedicated language feature, our fluent API designs use the host language compiler to enforce the protocol by ensuring that protocol violations result in type errors.

Language workbenches \cite{Fowler:language:workbenches,Erdweg:2013} such as Xtext \cite{Eysholdt:2010}, MPS \cite{Voelter:2012}, and MetaEdit+ \cite{Kelly:1996} are development environments for creating DSL development tools.
The Spoofax language workbench, for instance, compiles DSL specifications into a full-fledged DSL IDE \cite{Kats:2010}.
SugarJ \cite{Erdweg:2011}, Polyglot \cite{Nystrom:2003}, and Racket \cite{Felleisen:2015} are examples of extensible programming languages in which the syntax of the language can be extended to incorporate DSLs.
By embedding the DSL as a fluent API, our techniques work out of the box in statically-typed functional languages.

Programming languages with powerful compile-time metaprogramming support provide a direct solution to DSL embedding \cite{Czarnecki:2004}.
MetaML \cite{Sheard:1999}, MetaOCaml \cite{Calcagno:2003,Kiselyov:2014}, and Template Haskell \cite{Sheard:2002} extend existing programming languages with macro systems that enable DSL code generation at compile time.
Evidently, Haskell is expressive enough to embed DSLs even without templates.
Haskell supports ad hoc polymorphism through type classes \cite{Wadler:1989}, shown to be useful for DSL embedding \cite{Augustsson:2008,Hudak:1998}.
EDSLs often employ Haskell's \texttt{do}-statement and monadic comprehensions \cite{Gill:2014,Bracker:2014}, especially in parser combinator libraries \cite{Leijen:2000,Hutton:1998,Leijen:2001,Florijn:1995}.
In contrast to previous approaches, functional fluent APIs do not rely on any macro system, language extension, or special control structure.
Some of the Haskell EDSL techniques are ad hoc, whereas our fluent API solutions support well-defined classes of DSLs, namely all regular and deterministic context-free languages.

Existing fluent API designs generally do not apply to functional languages because of their lack of function overloading.
Enhancing functional languages with ad hoc polymorphism, e.g., as proposed by \citet{Bourdoncle:1997}, \citet{Shields:2001}, and \citet{Neubauer:2002}, may enable these designs to work in functional settings.
\citet{Yamazaki:2019} managed to encode fluent APIs in Haskell by using typeclasses and a few language extensions.
Their work and other modern fluent API designs are discussed in \cref{section:preliminaries}.
Our fluent APIs overcome the overloading problem by avoiding it altogether in favor of rudimentary functional features.

Indeed, we chose SML as the main language used throughout the paper for its relative simplicity.
Noting SML's rich module system \cite{MacQueen:1984}, we limit ourselves to core language constructs \cite{Harper:1986}.
Fluent API in SML is not a new concept:
\citet{Kamin:1997} presented a fluent API-like EDSL eight years before the term was even coined \cite{Fowler:2005}.

\subsection{Contributions}

We present different algorithms for encoding fluent APIs in functional languages.
Our fluent API designs support all regular and deterministic context-free API protocols and DSLs and can be implemented in most statically-typed functional languages.
This work is accompanied by an SML implementation of key algorithms.
\ifdefined\FullVersion
The appendix includes a series of experiments conducted using the implementation.
\else
The full paper \cite{Roth:2023} describes a series of experiments conducted using the implementation.
\fi

\paragraph*{Outline}
The rest of the paper is organized as follows.
\Cref{section:preliminaries} covers preliminary knowledge in automata theory and object-oriented fluent APIs.
The next three sections present three methods for encoding regular protocols as functional fluent APIs:
\emph{bit shuffling} in \cref{section:shuffling}, \emph{Church Booleans} in \cref{section:church}, and \emph{tabulation} in \cref{section:tabulation}.
In \cref{section:beyond} we show how to extend these techniques to non-regular protocols.
\Cref{section:zz} concludes with a discussion on functional fluent APIs and a comparison of the various encoding methods.

\section{Preliminaries}
\label{section:preliminaries}
\subsection{Automata and Formal Languages}

We assume that the reader is familiar with the structure and operation of deterministic finite state machines (FSMs) and pushdown automata (DPDA).
Here we briefly review some definitions;
for the full theoretical background, we recommend \citet{Hopcroft:2007}.

FSMs recognize the class of regular languages.
An FSM $M$ is a tuple $M=\langle Q, \Sigma, q_0, \delta, F \rangle$, where $Q$ is a finite set of states, $\Sigma$ is a finite set of input letters, $q_0 \in Q$ is the initial state, $\delta:Q\times\Sigma\rightarrow Q$ is the (partial) transition function, and $F \subseteq Q$ is the set of accepting states.
Let $\delta_\sigma:Q\rightarrow Q$ be $\delta$'s projection on $\sigma$, i.e., $\delta(q, \sigma)=p \Leftrightarrow \delta_\sigma(q)=p$.

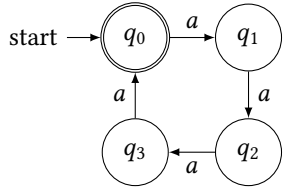
\begin{figure}[ht]
    \centering
    \begin{tikzpicture}[node distance=10ex]
        \node[state,initial,accepting] (q0) {$q_0$};
        \node[state,right of=q0] (q1) {$q_1$};
        \node[state,below of=q1] (q2) {$q_2$};
        \node[state,below of=q0] (q3) {$q_3$};
        \draw[->]
            (q0) edge[above] node{$a$} (q1)
            (q1) edge[right] node{$a$} (q2)
            (q2) edge[below] node{$a$} (q3)
            (q3) edge[left] node{$a$} (q0);
    \end{tikzpicture}
    \caption{An FSM accepting the language $(a^4)^*$}
    \label{figure:fsm:a4}
\end{figure}

The run of word $w=\sigma_1\sigma_2\ldots\sigma_n \in \Sigma^*$ on FSM $M$ is a sequence of states $\rho=p_0p_1\cdots p_n$ such that $p_0=q_0$ and for every $i=1,\ldots,n$, $\delta(p_{i-1}, \sigma_i) = p_i$.
We say that $M$ accepts $w$ if and only if $p_n \in F$; otherwise $M$ rejects $w$.
The language of $M$, $L(M)$ is the set of words $M$ accepts.
Consider, for example, the FSM depicted in \cref{figure:fsm:a4} in graph form.
This FSM recognizes the language $(a^4)^*$ containing all sequences of the letter $a$ whose length is a multiple of four.

A DPDA is an FSM that uses an auxiliary pushdown store
of stack symbols drawn from a finite set $\Gamma$.
The DPDA transition function is \[
    \delta:Q\times(\Sigma\cup\{\varepsilon\})\times\Gamma\rightarrow Q\times\Gamma^*
\]
In each computation step, the automaton reads symbol $\gamma \in \Gamma$ on the stack and replaces it with a (possibly empty) sequence of symbols $\boldsymbol\gamma \in \Gamma^*$.
If the the automaton tries to read from an empty stack, it gets stuck and rejects the input word.
A DPDA transition can consume an input letter $\sigma \in \Sigma$ or $\varepsilon$, the empty word.
These $\varepsilon$-transitions are used to modify the stack between input reads.
A DPDA without $\varepsilon$-transitions is called a real-time DPDA, as it makes exactly one operation for each input letter.
DPDAs recognize the class of deterministic context-free languages (DCFLs), a proper super-set of regular languages.
For instance, a DPDA can recognize the Dyck language of balanced brackets, but an FSM cannot.

\subsection{Encoding Automata as Fluent APIs in Object-Oriented Programming Languages}
\label{section:fluent}

The traditional study of compiler design \cite{Aho:1977} provides many parser generation techniques that take formal language specifications and convert them into parsers.
Similarly, the academic study of fluent APIs seeks DSL embedding techniques that encode DSL specifications (usually in automaton form) as fluent APIs.
We now review some of the existing fluent API encoding methods, designed primarily for object-oriented host languages.

Encoding an FSM as a fluent API is easy.
First, create class \inline{q} for each state $q \in Q$.
Then, for each transition $\delta(q, \sigma)=p$, add the method \inline{$\sigma$:q$\rightarrow$p} to class \inline{q}.
This method receives an (implicit) parameter of type \inline{q} and returns an instance of class \inline{p}.
Finally, add a special method \inline[literate=*{D}{\textdollar}{1}]{D} only to classes of accepting states $q_F \in F$ (its return type does not matter).
A fluent chain of API calls (chain for short) starts with an initial variable \inline{\_\_} of type $q_0$, followed by a sequence of API calls $\sigma_1,\sigma_2,\ldots,\sigma_n\in\Sigma$, and ends with a call to \inline[literate=*{D}{\textdollar}{1}]{D}:
\begin{equation}\label{eq:fluent}
    \text{\inline[language=java,literate=*{D}{\textdollar}{1}]{__().$\sigma_1$().$\sigma_2$().$\ldots$.$\sigma_n$().D()}}
\end{equation}

For example, the fluent API in \cref{listing:fluent:a4} encodes the language $(a^4)^*$ as defined by the FSM in \cref{figure:fsm:a4}.
The fluent API is followed by several chain examples in the form of \cref{eq:fluent}.
A given chain type checks if and only if the number of times it calls \inline{a()} is a multiple of four.
We say that the fluent API enforces (encodes, specifies, recognizes) the language $(a^4)^*$.

\begin{JAVA}[float=ht,language=java,caption={A Java fluent API encoding the language $(a^4)^*$},label={listing:fluent:a4}]
interface q0 { q1 a(); /* $\color{comment}\delta(q_0,a)=q_1$ */ void \$(); /* $\color{comment}q_0 \in F$ */ }
interface q1 { q2 a(); /* $\color{comment}\delta(q_1,a)=q_2$ */ }
interface q2 { q3 a(); /* $\color{comment}\delta(q_2,a)=q_3$ */ }
interface q3 { q0 a(); /* $\color{comment}\delta(q_3,a)=q_0$ */ }
q0 __ = null; // /* $\color{comment}q_0$ is the initial state
/* example chains */ {
    __.\$(); // compiles, $\varepsilon \in L((a^4)^*)$
    __.a().a().a().a().\$(); // compiles, $a^4 \in L((a^4)^*)$
    __.a().a().a().\$(); // does not compile, $a^3 \not\in L((a^4)^*)$
}
\end{JAVA}

So, how does it work?
The chain \cref{eq:fluent}
essentially simulates the FSM run $\rho=p_0p_1\cdots p_n$ of the word $w=\sigma_1\sigma_2\ldots\sigma_n$ encoded in the chain.
That is, the return type of \inline{$\sigma_i$()} in \cref{eq:fluent}, $i=1,2,\ldots,n$ is $p_i$.
Starting with the initial variable \inline{\_\_} of type $q_0$, the initial state, adding a call to \inline{$\sigma_i$()} yields the type $\delta(p_{i-1},\sigma_i)=p_i$.
The call \inline{$\sigma_n$()} returns an instance of class $p_n$, so the next call, to the terminal function \inline[literate=*{D}{\textdollar}{1}]{D()}, compiles if and only if $p_n \in F$, which is exactly the FSM membership condition.

Notice that method \inline{a} in \cref{listing:fluent:a4} is overloaded, since it has four different signatures: \inline{a:q0$\rightarrow$q1}, \inline{a:q1$\rightarrow$q2}, \inline{a:q2$\rightarrow$q3},
and \inline{a:q3$\rightarrow$q0}.
In general, this construction induces overloading whenever letter $\sigma \in \Sigma$ appears in more than one FSM transition.

Although challenging, it is possible to encode non-regular languages as fluent APIs \cite{Gil:2016,Nakamaru:2017,Xu:2010,Yamazaki:2019}.
We focus on the ``tree encoding'' of DPDAs introduced by \citet{Gil:2019}.
Gil and Roth presented an algorithm for converting a DPDA $A$ into a tree encoding that can be written as a fluent API.
The resulting fluent APIs look very similar to what we saw in \cref{listing:fluent:a4}, except that they use parametric polymorphism (generics, type parameters).
\Cref{listing:fluent:dyck} demonstrates how generics are used in the construction of a fluent API from tree encoding.
The listing contains a generic fluent API that recognizes the Dyck language---a well-known DCFL.
The chain examples that follow the fluent API compile if and only if they encode balanced left (\inline{L}) and right (\inline{R}) brackets.
The fluent API works by encoding the number of open brackets as a Peano number \inline{s<s<$\ldots$s<z>$\ldots$>>} in the chain's return types.

\begin{JAVA}[float=ht,language=java,caption={A Java fluent API encoding the Dyck language of balanced \texttt{L}/\texttt{R} brackets},label={listing:fluent:dyck}]
interface z { s<z> L(); void \$(); }
interface s<x> { s<s<x>> L(); x R(); }
z __ = null; // we only care about type checking at the moment
/* example chains */ {
    __.L().L().L().R().R().R().\$(); // compiles, the sequence ((())) is balanced
    __.L().L().R().L().R().R().\$(); // compiles, the sequence (()()) is balanced
    __.L().R().R().\$(); // does not compile, the sequence ()) is not balanced
    __.L().L().R().\$(); // does not compile, the sequence (() is not balanced
}
\end{JAVA}

We see Gil and Roth's tree encoding algorithm as a black box.
We only care about the structure of the fluent APIs it yields, as demonstrated in \cref{listing:fluent:dyck}.
In general, the tree encoding algorithm results in classes \inline{c0}, \inline{c1}, \ldots, \inline{cm} with zero or more type parameters, \inline{ci<x0, x1, $\ldots$, xn>}.
Functions' return types are general terms over the available classes \inline{c0}--\inline{cm} and parameters \inline{x0}--\inline{xn}, e.g., \inline{c1<c2<x3, c3>, x1>}.
Note that the tree encoding algorithm, as well as the encoding techniques proposed in the papers cited above, rely on overloading similarly to the FSM case.

\subsection{A Note on Typeclasses and Fluent APIs}

The object-oriented fluent API techniques mentioned above, including the more advanced designs of \citet{Gil:2019} and \citet{Yamazaki:2019}, can be adapted to functional programming languages by substituting typeclasses for function overloading.
Typeclasses were introduced by \citet{Wadler:1989} as a formalism for ad hoc polymorphism in Haskell and other functional programming languages.
Wadler and Blott defined the semantics of typeclasses by translating them to the standard Hindley-Milner type system.
This translation method was later called \emph{dictionary passing}.
\Cref{listing:dictionary:passing} demonstrates how to use dictionary passing to implement an ``overloaded'' function in SML.

\begin{JAVA}[float=ht,style=sml,caption={Dictionary passing in SML},label={listing:dictionary:passing}]
fun square (numD: { mul: 'a -> 'a -> 'a }) x = #mul numD x x (* an "overloaded" square function *)
fun mulInt (a: int) (b: int) = a * b (* multiplication of integers *)
val intD = { mul = mulInt } (* int dictionary *)
fun mulReal (a: real) (b: real) = a * b (* multiplication of real numbers *)
val realD = { mul = mulReal }; (* real dictionary *)
square intD 2; (* $\color{comment}\Rightarrow$ 4: int *)
square realD 3.0; (* $\color{comment}\Rightarrow$ 9.0: real *)
\end{JAVA}

Function \inline{square} in \cref{listing:dictionary:passing} accepts integers and real numbers.
We tell \inline{square} how to multiply numbers by passing it a dictionary \inline{numD} that points to the appropriate multiplication method: \inline{intD} for integers and \inline{realD} for real numbers.

Although dictionary passing is semantically equivalent to typeclasses, it is not a viable alternative for function overloading.
Matching each argument type with a dictionary (\inline{intD}, \inline{realD}) takes the same effort from the programmer as having multiple functions (\inline{squareInt}, \inline{squareReal}).
In Wadler and Blott’s original design, the compiler infers the dictionary from the argument type, so the programmer doesn't have to specify it explicitly.

\section{Bit Shuffling}
\label{section:shuffling}
The fluent API encoding techniques described in \cref{section:fluent} use function overloading and, therefore, cannot be used in functional languages such as SML.
This section presents \emph{bit shuffling}, the first of our three alternative methods for encoding FSMs as functional fluent APIs.
This method is unique in that it does not rely on unification, resulting in relatively fast compilation even when enforcing highly complex API protocols.

The crux of the new design is seeing the FSM states as binary vectors and the FSM transitions as a collection of Boolean functions over these vectors.
The current FSM state is encoded using a tuple of types \inline{f} and \inline{t} that represent the bits 0 and 1.
Each API function applies a certain transformation onto this tuple by \emph{shuffling} the bits comprising it:
\[
    \inline{(f, t, f, t)} \implies \inline{(t, f, f, t)}
\]
If the shuffling is done correctly, the resulting tuple encodes the subsequent FSM state.

We start by constructing a fluent API for the language $(a^4)^*$, as defined by the FSM in \cref{figure:fsm:a4} (this will be our running example).
Then, we generalize the construction steps for any FSM and assemble them into \cref{algorithm:shuffle}.

First, however, we define some notations.
A Boolean bit $b$ is a member of $\{0, 1\}$.
We also denote Boolean bits as \emph{false}, \inline{f}, and \inline{F} or \emph{true}, \inline{t}, and \inline{T}.
A Boolean vector $v$ of length $n$ is a sequence of $n$ Boolean bits, $v \in \{0, 1\}^n$.
We write $v$ as $\langle b_n, \ldots, b_2, b_1 \rangle$ (indexed from the right to the left) or as $b_n\cdots b_2b_1$ for short.
An $n$-ary Boolean function $f:\{0, 1\}^n\rightarrow\{0, 1\}$ accepts a Boolean vector of length $n$ and returns one Boolean bit.
An $n$-ary Boolean vector function $\boldsymbol{f}:\{0, 1\}^n\rightarrow\{0, 1\}^n$ accepts a Boolean vector of length $n$ and returns a vector of the same length.
A Boolean vector function can be decomposed into $n$ Boolean functions $\boldsymbol{f}=\langle f_n, \ldots, f_2, f_1 \rangle$ such that $f_i$ computes the $i$\hh bit of $\boldsymbol{f}$, \[
    \boldsymbol{f}(v) = \langle f_n(v), \ldots, f_2(v), f_1(v) \rangle
\]
The composition of Boolean function $g$ and vector function $\boldsymbol{f}$ is defined by
\begin{equation*}\label{eq:composition}
    (g \circ \boldsymbol{f})(v)=g(\boldsymbol{f}(v))
\end{equation*}
Finally, the composition of Boolean vector functions $\boldsymbol{g}$ and $\boldsymbol{f}$ is defined as \[
    \boldsymbol{g} \circ \boldsymbol{f} = \langle g_n \circ \boldsymbol{f}, \ldots, g_2 \circ \boldsymbol{f}, g_1 \circ \boldsymbol{f} \rangle
\]
Note that $(\boldsymbol{g} \circ \boldsymbol{f})(v)=\boldsymbol{g}(\boldsymbol{f}(v))$.

We are now ready to describe the FSM encoding steps.

\subsection{Step I: Binary Encoding of States and Transitions}
\label{section:shuffling:1}

Encode the automaton states $Q=\{q_0, q_1, q_2, q_3\}$ as Boolean vectors of length $\log(|Q|)=2$ using an arbitrary injective mapping $\beta: Q\rightarrow\{0, 1\}^2$:
\begin{equation*}\label{eq:beta}
    \begin{aligned}
        \beta(q_0)&=00 \\
        \beta(q_1)&=01 \\
        \beta(q_2)&=10 \\
        \beta(q_3)&=11
    \end{aligned}
\end{equation*}
Recall that $\delta_a$ is the projection of the transition function $\delta$ on letter $\sigma=a$.
Now, construct a Boolean vector function $\boldsymbol{f}^a$ such that $\boldsymbol{f}^a(\beta(q))=\beta(p) \Leftrightarrow \delta_a(q)=p$:
\begin{equation}\label{eq:f}
    \begin{aligned}
        \boldsymbol{f}^a(00)&=01 \\
        \boldsymbol{f}^a(01)&=10 \\
        \boldsymbol{f}^a(10)&=11 \\
        \boldsymbol{f}^a(11)&=00
    \end{aligned}
\end{equation}

Let $\boldsymbol{f}^a=\langle f^a_2, f^a_1 \rangle$.
Notice that $f^a_2$ is exclusive or (XOR), $f^a_2=\oplus$ and $f^a_1$ is negation (NOT) applied to the LSB, $f^a_1=\neg_1$, so we can write \[
    \boldsymbol{f}^a=\langle \oplus, \neg_1 \rangle
\]
Observe that we can describe the FSM run on word $w=a^n$ using a sequence of $n$ applications of $\boldsymbol{f}^a$:
\begin{equation*}\label{eq:shuffling:run}
    (\boldsymbol{f}^a \circ \boldsymbol{f}^a \circ \ldots \circ \boldsymbol{f}^a)(\beta(q_0))=\beta(p_n)
\end{equation*}
where $p_n$ is the last state in $w$'s run
(this observation is proved by a simple induction on $n$).
We can say that $\boldsymbol{f}^a$ consolidates the four FSM transitions comprising $\delta_a$ into a single Boolean vector function.

Intuitively, the consolidation of FSM transitions into a single function helps us overcome the overloading problem, which arose in the first place because of the multiplicity of transitions.
It is not enough, however, to implement $\boldsymbol{f}^a$ as a function on Boolean vectors.
Let
\begin{equation}\label{eq:shuffling:chain}
    \text{\inline[literate=*{D}{\textdollar}{1}{C}{\textasciicircum}{1}]{CC a a\ $\ldots$ a D}}
\end{equation}
be a functional fluent chain, with \inline[literate=*{C}{\textasciicircum}{1}]{CC} replacing \inline{\_\_} as the initial variable (now a function).
We want the chain \cref{eq:shuffling:chain} to compile if and only if the number of times \inline{a} is called is a multiple of four, i.e., if the chain encodes a word $w \in L((a^4)^*)$.
Yet, if we implement the fluent API as a product of Boolean vector functions, say,
\begin{JAVA}[style=sml]
val ^^ = (false, false) (* $\color{comment}\beta(q_0)=00$ *)
fun a (b2, b1) = (b2 xor b1, not b1) (* $\color{comment}\boldsymbol{f}^a=\langle\oplus,\neg_1\rangle$ *)
fun \$ (b2, b1) = if not b2 andalso not b1 then () else raise Fail "!" (* $\color{comment}q \in F \Leftrightarrow \beta(q)=00$ *)
\end{JAVA}
the bits \inline{b2} and \inline{b1} would be evaluated at run time, but the fluent API must inspect them at compile time in order to fail compilation (if needed).
To make the Boolean values available at compile time, we implement them as constants of different types:
\begin{JAVA}[style=sml]
datatype t = T (* 1, true Boolean *)
     and f = F (* 0, false Boolean *)
\end{JAVA}
The return type of \inline[style=sml]{fun a (b2, b1)}, now however, depends on the specific types of its arguments, which brings us back to the overloading problem.
We address this issue in Step II.

\subsection{Step II: Direct Encoding of Boolean Function Evaluations}
\label{section:shuffling:2}

Consider the following Boolean functions:
\begin{equation}\label{eq:gs}
    \begin{aligned}
        g^1(b_2b_1) &= b_2 \wedge b_1 \\
        g^2(b_2b_1) &= b_2 \wedge \neg b_1 \\
        g^4(b_2b_1) &= \neg b_2 \wedge b_1 \\
        g^8(b_2b_1) &= \neg b_2 \wedge \neg b_1
    \end{aligned}
\end{equation}
Function $g^8$ can be used to check our FSM's acceptance condition, since \[
    q \in F \iff q=q_0 \iff g^8(\beta(q))=1
\]
In the bit shuffling method, the terminal function \inline[literate=*{D}{\textdollar}{1}]{D} does not evaluate $g^8$ by itself, but instead receives its evaluation as an argument \inline{g8} (to replace \inline{b2} and \inline{b1}).
Then, it is a simple matter to check that the type of \inline{g8} is \inline{t}, i.e., that $g^8(\beta(p_n))=1$.

Let $w=a^n$ be the word encoded by the fluent chain \cref{eq:shuffling:chain}, and let $\rho=p_0p_1\cdots p_n$ be its run.
The initial variable \inline[literate=*{C}{\textasciicircum}{1}]{CC} assigns \inline{g8=T}, since \[
    g^8(\beta(p_0))=g^8(\beta(q_0))=1
\]
The API function \inline{a} should receive in parameter \inline{g8} the evaluation of $g^8(\beta(p_i))$ and return the evaluation of $g^8(\beta(p_{i+1}))$.
Since $\boldsymbol{f}^a(\beta(q))=\beta(p) \Leftrightarrow \delta_a(q)=p$ (see \cref{eq:f}), $g^8(\beta(p_{i+1}))$ can be computed as follows: \[
    g^8(\beta(p_{i+1})) =
    g^8(\beta(\delta_a(p_i))) =
    g^8(\boldsymbol{f}^a(\beta(p_i))) = \ldots
\]
Let $\beta(p_i)=b_2b_1$. We continue: \[
    \ldots =
    g^8(\boldsymbol{f}^a(b_2b_1)) =
    g^8(\langle b_2 \oplus b_1, \neg b_1\rangle) =
    \neg(b_2 \oplus b_1) \wedge b_1 =
    b_2 \wedge b_1 =
    g^1(b_2b_1) =
    g^1(\beta(p_i))
\]
(Function $g^1$ is defined in \cref{eq:gs}.)
Overall, we get \begin{equation}\label{eq:g8}
    g^8(\beta(p_{i+1}))=(g^8 \circ \boldsymbol{f}^a)(\beta(p_i))=g^1(p_i)
\end{equation}
i.e., the evaluation of $g^8$ on the next state $p_{i+1}$ is equal to applying the composition $g^8 \circ \boldsymbol{f}^a$ to the current state $p_i$, which in turn equals the evaluation of another function $g^1$ on $p_i$.

We let \inline{a} receive the value of $g^1(\beta(p_i))$ in yet another parameter \inline{g1}.
We then continue recursively to compute $g^1(\beta(p_{i+1}))$, in a similar way to how $g^8(\beta(p_{i+1}))$ was computed:
\begin{equation}\label{eq:g1:g2:g4}
    \begin{aligned}
        g^1(\beta(p_{i+1}))=(g^1\circ \boldsymbol{f}^a)(\beta(p_i))=g^2(\beta(p_i)) \\
        g^2(\beta(p_{i+1}))=(g^2\circ \boldsymbol{f}^a)(\beta(p_i))=g^4(\beta(p_i)) \\
        g^4(\beta(p_{i+1}))=(g^4\circ \boldsymbol{f}^a)(\beta(p_i))=g^8(\beta(p_i))
    \end{aligned}
\end{equation}
The recursion ends on $g^8(\beta(p_i))$, since this value is already given in parameter \inline{g8}.
Let $\boldsymbol{g}=\langle g^1, g^2, g^4, g^8 \rangle$ (the ordering of functions in $\boldsymbol{g}$ does not matter).
\Cref{eq:g8,eq:g1:g2:g4} can be summarized as follows:
\begin{equation}\label{eq:g}
    \boldsymbol{g} \circ \boldsymbol{f}^a = \langle g^2, g^4, g^8, g^1 \rangle
\end{equation}

Thus, function \inline{a} receives a quadruple \inline{(g1, g2, g4, g8)} representing the four evaluations of $g^j(\beta(p_i))$, $j=1, 2, 4, 8$, and returns the evaluations of $g^j(\beta(p_{i+1}))$ by shuffling the quadruple into \inline{(g2, g4, g8, g1)} following \cref{eq:g}.

The resulting fluent API is shown in \cref{figure:fsm:a4}.
The addition of parameter \inline[style=sml]{f'} to functions \inline[literate=*{C}{\textasciicircum}{1}]{CC} and \inline{a} allows us to call them in fluent form, i.e., from the left to the right as in \cref{eq:shuffling:chain}.

\begin{JAVA}[float=ht,style=sml,caption={A shuffling SML fluent API for the FSM of \cref{figure:fsm:a4}},label={listing:shuffle:a4}]
datatype t = T (* 1, true Boolean *)
     and f = F (* 0, false Boolean *)
fun ^^ f' = f' (F, F, F, T) (* $\color{comment}\langle g_1, g_2, g_4, g_8 \rangle(\beta(q_0))=\langle g_1(00), g_2(00), g_4(00), g_8(00) \rangle=0001$ *)
fun a (g1, g2, g4, g8) f' = f' (g2, g4, g8, g1) (* $\color{comment}\langle g_1, g_2, g_4, g_8 \rangle \circ \boldsymbol{f}^a = \langle g_2, g_4, g_8, g_1 \rangle$ *)
fun \$ (_, _, _, T) = (); (* $\color{comment}q \in F \Leftrightarrow g_8(\beta(q))=1$ *)
(* chain examples: *)
^^ \$; (* OK, $\color{comment}\varepsilon\in L((a^4)^*)$ *)
^^ a a a a \$; (* OK, $\color{comment}a^4\in L((a^4)^*)$ *)
^^ a a a a a a a a \$; (* OK, $\color{comment}a^8\in L((a^4)^*)$ *)
^^ a \$; (* error, $\color{comment}a^1\not\in L((a^4)^*)$ *)
^^ a a \$; (* error, $\color{comment}a^2\not\in L((a^4)^*)$ *)
^^ a a a a a \$; (* error, $\color{comment}a^5\not\in L((a^4)^*)$ *)
\end{JAVA}

The fluent API in \cref{figure:fsm:a4} is followed by a few fluent chain examples.
Notice that a chain compiles if and only if the number of \inline{a} calls is a multiple of four, i.e., if the chain encodes $w \in L((a^4)^*)$.

\subsection{The Bit Shuffling Encoding Algorithm}
\label{section:shuffling:algorithm}

We now generalize the encoding steps described in \cref{section:shuffling:1,section:shuffling:2} and present an algorithm for any FSM $M$.
FSMs are, in general, more complicated than our example FSM, so we need to add some minor steps to the encoding process.
First off, notice that our example FSM, depicted in \cref{figure:fsm:a4}, has a total transition function $\delta$, i.e., every state has an outgoing edge for each input letter, and the number of states $|Q|=4$ is a power of two.
These properties make the Boolean vector function $\boldsymbol{f}^a$ well-defined, i.e., $\boldsymbol{f}^a(v)$ is defined for any Boolean vector of length $n=\log(|Q|)$.
Given an FSM $M$, we make its transition function total by adding a non-accepting \emph{sink} state, to which the missing transitions are directed.
The sink state loops for any letter $\sigma\in\Sigma$.
We then complete the number of states to a power of two by, again, adding sink states.
These modifications result in automaton $M'=\langle Q', \Sigma, q_0, F, \delta'\rangle$ that accepts the same language as $M$, $L(M')=L(M)$.

Let $n=\log(|Q'|)$, $n\in\mathbb{N}$ be the number of bits in mapping $\beta$.
Our example fluent API required the evaluation of four Boolean functions, $g^1$, $g^2$, $g^4$, and $g^8$.
Instead of calculating which $g^i$ functions are required for $M'$, we just take all the Boolean functions on $n$ bits, $\boldsymbol{g}^n=\langle g^1, g^2, g^3, \ldots\rangle$ (there is a finite number of such functions).
In addition, notice that the example language $(a^4)^*$ uses only a single letter $a$, but we want to support alphabets $\Sigma$ of any finite size.
Given alphabet $\Sigma$ of arbitrary size, we simply repeat the encoding steps described above for each letter $\sigma \in \Sigma$, i.e., compute $\boldsymbol{f}^\sigma$ and $\boldsymbol{g}^n \circ \boldsymbol{f}^\sigma$ and encode the result in a function \inline{$\sigma$}.

Finally, the example FSM specifies a single accepting state $F=\{q_0\}$, but FSMs in general may have multiple accepting states.
Therefore, we need to find a Boolean function $g^F$ such that $g^F(\beta(q))=1 \Leftrightarrow q \in F$.
There must be (exactly one) such function in $\boldsymbol{g}^n$.
The terminal function simply checks that the type of parameter \inline{gF}, holding the evaluation of $g^F$, is \inline{t}.

The complete bit shuffling algorithm is presented in \cref{algorithm:shuffle}.

\begin{algorithm}
    \caption{The bit shuffling algorithm for encoding an FSM as an SML fluent API}
    \label{algorithm:shuffle}
    \centering
    \fcolorbox{black!25}{white}{
    \begin{minipage}{.91\textwidth}
    \begin{algorithmic}[1]
        \INPUT FSM $M=\langle Q, \Sigma, q_0, F, \delta\rangle$
        \PRINT{\inline[style=sml]{datatype t = T and f = F}}
        \STATE{Construct FSM $M'=\langle Q', \Sigma, q_0, F, \delta'\rangle$ from $M$ by making the transition function $\delta$ total and adding states until $|Q'|$ is a power of two (both using sink states)}
        \LET{$n$}{$\log(|Q'|)$}
        \LET{$\boldsymbol{g}^n$}{all Boolean functions on $n$ bits}
        \LET{$\beta$}{an injective mapping $\beta:Q'\rightarrow \{0, 1\}^n$}
        \PRINT{\inline[style=sml,literate=*{C}{\textasciicircum}{1}]{fun CC f' = f'\ $\boldsymbol{g}^n(\beta(q_0))$}}
        \FORALL{$\sigma\in\Sigma$}
            \LET{$\boldsymbol{f}^\sigma$}{a mapping $\{0, 1\}^n\rightarrow\{0, 1\}^n$ such that $\boldsymbol{f}^\sigma(\beta(q))=\beta(p) \Leftrightarrow \delta_\sigma(q)=p$}
            \PRINT{\inline[style=sml]{fun\ $\sigma$\ $\boldsymbol{g}^n$ f' = f'\ $\boldsymbol{g}^n \circ \boldsymbol{f}^\sigma$}}
        \ENDFOR
        \LET{$g^F$}{the (one and only) function in $\boldsymbol{g}^n$ such that $g^F(\beta(q))=1 \Leftrightarrow q \in F$}
        \PRINT{\inline[style=sml,literate=*{D}{\textdollar}{1}{E}{\hbox to 1em{.\hss.\hss.}}{1}]{fun D (_, E, _, T, _, E, _) = ()} where \inline{T} appears in the position of $g^F$ in $\boldsymbol{g}^n$}
    \end{algorithmic}
    \end{minipage}}
\end{algorithm}

Some of the code excerpts printed in \cref{algorithm:shuffle} contain mathematical symbols---we now explain how these are rendered as text:
\begin{itemize}
    \item A Boolean bit $b$ is rendered as \inline{T} if $b=1$ or as \inline{F} if $b=0$.
    \item A Boolean vector $v=\langle b_n, \ldots, b_2, b_1 \rangle$ is rendered as a tuple \inline[literate=*{E}{\hbox to 1em{.\hss.\hss.}}{1}]{($b_n$, E,\ $b_2$,\ $b_1$)}.
    \item A Boolean function $g$ is rendered as \inline{g}, i.e., we ignore its intent and print its name.
    \item A Boolean vector function $\boldsymbol{g}=\langle g_1, g_2, \ldots, g_n \rangle$ is rendered as a tuple \inline[literate=*{E}{\hbox to 1em{.\hss.\hss.}}{1}]{($g_1$,\ $g_2$, E,\ $g_n$)}.
\end{itemize}

\subsection{Encoding Improvements}
\label{section:shuffling:improvements}

We now propose two optional improvements of the bit shuffling algorithm as it is described in \cref{algorithm:shuffle}.
First, notice that \cref{algorithm:shuffle} computes all the binary functions on $n=\log(|Q'|)$ bits.
Overall, there are \[
    2^{2^n}=2^{|Q'|}\le 2^{2|Q|}=4^{|Q|}
\] such functions, so the run time of the algorithm, as well as the size of the code it prints, are exponential in the size of the FSM.
Nevertheless, our example fluent API (\cref{listing:shuffle:a4}) uses only four of these functions (out of 16), indicating that it might be possible to get rid of at least some functions.
To determine which functions are necessary and which can be omitted, we run the recursive procedure described in \cref{section:shuffling:2}:
Start with $\boldsymbol{g}^n=\langle g^F\rangle$, and recursively add function $g$ if there exists $\sigma \in \Sigma$ and $g' \in \boldsymbol{g}^n$ such that $g' \circ \boldsymbol{f}^\sigma = g$.
From our experience, this construction of $\boldsymbol{g}^n$ reduces its size significantly for ``reasonable'' FSMs.
Determining the exact degree to which $\boldsymbol{g}^n$ is reduced and, specifically, whether or not its sizes stays exponential in $|Q|$, is left as an open problem.

The second improvement relates to the fluent API user experience.
The fluent API printed by \cref{algorithm:shuffle} allows any sequence of API calls to be made and checks whether or not it makes sense only in the terminal method \inline[literate=*{D}{\textdollar}{1}]{D}.
Therefore, the programmer using the fluent API is not alerted when they call a wrong API method, and when they do (on calling \inline[literate=*{D}{\textdollar}{1}]{D}), they are not told where the mistake is located.
Let
\begin{equation}\label{eq:fluent:partial}
    \text{\inline[literate=*{C}{\textasciicircum}{1}]{CC\ $\sigma_1$\ $\sigma_2$\ $\ldots$\ $\sigma_n$}}
\end{equation}
be a partial fluent chain encoding $w=\sigma_1\sigma_2\cdots\sigma_n$, and let $\rho=p_0p_1\cdots p_n$ be $w$'s run on the FSM.
To make our fluent API more user friendly, we ensure that a call to $\sigma_{n+1} \in \Sigma$ after \cref{eq:fluent:partial} will not compile if the resulting chain cannot be completed correctly by some sequence of API calls.
We call this property of fluent APIs \emph{early failure}.

Let $R \subseteq Q$ be the set of states from which an accepting state is reachable (when viewing the FSM in graph form).
Then, the call to $\sigma_{n+1}$ should compile if and only if \begin{equation}\label{eq:r}
    \delta(p_n, \sigma_{n+1})\in R
\end{equation}
We can check this condition using the Boolean function $g^R$ (similar to $g^F$ used by \inline[literate=*{D}{\textdollar}{1}]{D}) that holds \[
    g^R(\beta(q))=1 \iff q \in R
\]
We require that the evaluation of $g^R$, stored in parameter \inline{gR}, is always true throughout the chain to achieve early failure.
Let $\sigma\in\Sigma$ be an API function, and let \inline{gi} be the parameter it assigns in the position of \inline{gR} in the shuffled tuple on function $\sigma$'s right-hand side, i.e., $g^R(\beta(p_{j+1}))=g_i(\beta(p_j))$.
If $g^R$ evaluates to false after calling $\sigma$, then our encoding ensures that \inline{gi} is also false.
Function \inline{$\sigma$} should, accordingly, check that the type of \inline{gi} is \inline{t} by replacing the parameter \inline{gi} with \inline{T}, as done in function \inline[literate=*{D}{\textdollar}{1}]{D} for parameter \inline{gF}.

When the two improvements described above are applied together, the recursive construction of $\boldsymbol{g}^n$ should start with $\boldsymbol{g}^n=\langle g^F, g^R \rangle$ to ensure that the parameter \inline{gR} is present in the final vector.

\section{Church Booleans}
\label{section:church}
This section describes our fluent API encoding method based on the Church encoding of Booleans.
In contrast to bit shuffling, presented in \cref{section:shuffling}, the current technique relies on type inference and thus incurs longer compilation times.
Nevertheless, our technique demonstrates an application of the classic Church Booleans to API design which, we believe, is a fascinating concept whose full potential is beyond the scope of the current paper.

Recall that bit shuffling encodes FSM states as Boolean vectors and FSM transitions as functions over these vectors.
The same idea is at the base of our Church encoding.
But instead of pre-computing function evaluations, as done in bit shuffling, the new technique directly computes bit functions at compile time by replacing the true and false monotypes with the Church Booleans.
While Church Boolean circuits are usually applied in an untyped manner, they work mostly the same at the type level, except for a minor digression: duplicating a bit into two identical bits must be expressed explicitly using a fanout gate.

Before presenting the encoding, we review Church Booleans and how they can be implemented in SML:
\begin{JAVA}[style=sml]
fun T x y = x (* true, T : 'a$\color{comment}\rightarrow$'b$\color{comment}\rightarrow$'a *)
fun F x y = y (* false, F : 'a$\color{comment}\rightarrow$'b$\color{comment}\rightarrow$'b *)
\end{JAVA}
\emph{True} is a function that receives two arguments (Currying style) and returns the first;
\emph{False} also receives two parameters but returns the latter.
The type of true is \inline[style=sml]{T:'a->'b->'a} and the type of false is \inline[style=sml]{F:'a->'b->'b}.
These are the Church Boolean functions:\footnote{If you try to run this code using SML/NJ, you will get ``dummy types'' instead of plain \inline[style=sml]{'a} and \inline[style=sml]{'b} type parameters. We ignore dummy types in the current discussion.}
\begin{JAVA}[style=sml]
fun Not b = b F T
fun Or (b2, b1) = b2 T b1
fun And (b2, b1) = b2 b1 F;
Not T; (* false, 'a$\color{comment}\rightarrow$'b$\color{comment}\rightarrow$'b *)
Or (And (T, F), And (T, T)); (* true, 'a$\color{comment}\rightarrow$'b$\color{comment}\rightarrow$'a *)
\end{JAVA}

Can we use these simple Boolean functions to describe complex Boolean circuits?
The following attempt fails to compile:
\begin{JAVA}[style=sml]
fun Xor (b2, b1) = Or (And (b2, Not b1), And (Not b2, b1));
Xor (T, F); (* compilation error: circularity *)
\end{JAVA}
Evidently, the SML compiler fails to assign reasonable types to the parameters of function \inline{Xor}.
\citet{Mairson:2004} explains that function \inline{Xor} is problematic because it is non-linear, in the sense that it uses the bits \inline{b2} and \inline{b1} more than once in the same expression.
To restore linearity, Mairson proposes to explicitly implement a fanout gate \inline{Copy} that takes a single bit and returns two equivalent bits:
\begin{JAVA}[style=sml]
fun Pair x y z = z x y
fun Copy p = p (Pair T T) (Pair F F)
\end{JAVA}
Function \inline{Copy} is used this way: \[
    \text{\inline[style=sml]{Copy b (fn b1 => fn b2 =>\ $e$)}}
\]
\inline{Copy} evaluates the inner expression \inline{$e$} by assigning \inline{b}'s type (Boolean value) to \inline{b1} and \inline{b2}.
Therefore, \inline{$e$} can use two instances \inline{b1} and \inline{b2} of the same Boolean bit \inline{b}.
Let us re-implement \inline{Xor} using \inline{Copy}:
\begin{JAVA}[style=sml]
fun Xor (b2, b1) =
    Copy b2 (fn b2_1 => fn b2_2 =>
    Copy b1 (fn b1_1 => fn b1_2 =>
        Or (And (b2_1, Not b1_1), And (Not b2_2, b1_2))
    ));
Xor (T, F); (* true, 'a->'b->'a *)
Xor (F, F); (* false, 'a->'b->'b *)
\end{JAVA}

After learning how to compute complex Boolean circuits at compile time, we are ready to describe the Church encoding of fluent APIs.
Again, we start by encoding a fluent API for our example language $(a^4)^*$ and then generalize the construction for any FSM in \cref{algorithm:church}.
The first step of the encoding is identical to Step I of bit shuffling (described in \cref{section:shuffling:1}), so we can continue straight to the second step.

\subsection{Step II: Encoding FSM States as Vectors of Church Booleans}

In Step I (\cref{section:shuffling:1}) we constructed mapping $\beta: Q \rightarrow \{0, 1\}^2$ from states to bit vectors.
Function $\beta$ is used to encode the FSM states $Q$ as Church Boolean vectors:
\begin{JAVA}[style=sml]
val q0 = (F, F) (* $\color{comment}q_0=00$, q0 : ('a->'b->'b) * ('c->'d->'d) *)
val q1 = (F, T) (* $\color{comment}q_1=01$, q1 : ('a->'b->'b) * ('c->'d->'c) *)
val q2 = (T, F) (* $\color{comment}q_2=10$, q2 : ('a->'b->'a) * ('c->'d->'d) *)
val q3 = (T, T) (* $\color{comment}q_3=11$, q3 : ('a->'b->'a) * ('c->'d->'c) *)
\end{JAVA}

We also defined the vector function $\boldsymbol{f}^a:\{0, 1\}^2\rightarrow \{0, 1\}^2$ where \[
    \boldsymbol{f}^a(\beta(q))=\beta(p) \iff \delta_a(q)=p
\]
This function can be described as a composition of two Boolean functions $\boldsymbol{f}^a=\langle \oplus, \neg_1 \rangle$, i.e., \[
    \boldsymbol{f}^a(b_2b_1)=\langle b_2 \oplus b_1, \neg b_1\rangle
\]
Function $\boldsymbol{f}^a$ can be encoded in SML using Church Booleans and Mairson's \inline{Copy} function as follows:
\begin{JAVA}[style=sml]
fun a (b2, b1) = Copy b1 (fn b1_1 => fn b1_2 =>
    (Xor (b2, b1_1), Not b1_2)
)
\end{JAVA}
(for \inline{Not} and \inline{Xor} described above).
Let us see function \inline{a} in action:
\begin{JAVA}[style=sml]
a q0; (* q1, ('a->'b->'b)*('c->'d->'c)  *)
a q1; (* q2, ('a->'b->'a)*('c->'d->'d)  *)
a q2; (* q3, ('a->'b->'a)*('c->'d->'c)  *)
a q3; (* q0, ('a->'b->'b)*('c->'d->'d)  *)
\end{JAVA}
The type of \inline{a} applied on \inline{qi} is the type of \inline{qj} where $\delta_a(q_i)=q_j$.
Therefore, we can simulate an FSM run by recursively applying \inline{a} to \inline{q0}.
Let $w=a^n$ and let $p_n$ be the last state in $w$'s run.
Then, applying function \inline{a} to \inline{q0} $n$ times results in an expression of type $p_n$ (the inductive proof is simple).

The terminal function \inline[literate=*{D}{\textdollar}{1}]{D} should check that the last state in the run is accepting.
Recall that in our example, this condition is verified by \[
    g^8(b_2b_1) = \neg b_2 \wedge \neg b_1
\]
as $g^8(\beta(q))=1 \Leftrightarrow q \in F$.
Since function $g^8$ is linear, it can be encoded simply as \[
    \inline[style=sml]{fun g8 (b2, b1) = And (Not b2) (Not b1)}
\]
Function \inline[literate=*{D}{\textdollar}{1}]{D} applies $g^8$ to $b_2b_1$ and compels the result \inline{b} to be true:
\begin{JAVA}[style=sml]
fun \$ (b2, b1) = let val b = g8 (b2, b1) in (b 0 "") + 0 end
\end{JAVA}
The expression \inline{(b 0 "")} has type \inline[style=sml]{int} if \inline{b} is true or \inline[style=sml]{string} otherwise, so by adding \inline{+0} we force \inline{b} to be true, or else compilation will fail since a \inline[style=sml]{string} cannot be added to an \inline[style=sml]{int} in SML.

The full encoding of $(a^4)^*$ is shown in \cref{listing:church:a4}, together with a few example chains.

\begin{JAVA}[float=ht,style=sml,caption={A Church-encoded SML fluent API for the FSM of \cref{figure:fsm:a4}},label={listing:church:a4}]
fun T x y = x (* true, T : 'a->'b->'a *)
fun F x y = y (* false, F : 'a->'b->'b *)
fun Not b = b F T
fun Or (b2, b1) = b2 T b1
fun And (b2, b1) = b2 b1 F
fun Pair x y z = z x y
fun Copy p = p (Pair T T) (Pair F F)
fun Xor b1 b2 =
    Copy b2 (fn b2_1 => fn b2_2 =>
    Copy b1 (fn b1_1 => fn b1_2 =>
        Or (And (b1_1, Not b2_1), And (Not b1_2, b2_2))
    ))
fun ^^ f' = f' (F, F) (* $\color{comment}\beta(q_0)=00$ *)
fun a (b2, b1) f' = f' (Copy b1 (fn b1_1 => fn b1_2 =>
    (Xor b2 b1_1, Not b1_2)
)) (* $\color{comment}\boldsymbol{f}^a(b_2b_1)=\langle b_2 \oplus b_1, \neg b_1\rangle$ *)
fun g8 (b2, b1) = And (Not b2) (Not b1) (* $\color{comment}g^8(\beta(q))=1 \Leftrightarrow q=q_0$ *)
fun \$ (b2, b1) = let val b = g8 (b2, b1) in (b 0 "") + 0 end; (* accepts only $\color{comment}b_2b_1=00=\beta(q_0)$ *)
(* chain examples: *)
^^ \$; (* OK, $\color{comment}\varepsilon\in L((a^4)^*)$ *)
^^ a a a a \$; (* OK, $\color{comment}a^4\in L((a^4)^*)$ *)
^^ a \$; (* error, $\color{comment}a^1\not\in L((a^4)^*)$ *)
^^ a a \$; (* error, $\color{comment}a^2\not\in L((a^4)^*)$ *)
^^ a a a a a \$; (* error, $\color{comment}a^5\not\in L((a^4)^*)$ *)
\end{JAVA}

\subsection{The Church Boolean Encoding Algorithm}

As in bit shuffling, the general Church Boolean encoding algorithm starts by converting the input FSM $M$ into an equivalent FSM $M'$ with a total transition function and whose number of states is a power of two.
The other generalization steps (to support multiple letters and multiple accepting states) are the same as in bit shuffling (see \cref{section:shuffling:algorithm}).
The full algorithm is shown in \cref{algorithm:church}.

\begin{algorithm}
    \caption{The Church Boolean algorithm for encoding an FSM as an SML fluent API}
    \label{algorithm:church}
    \centering
    \fcolorbox{black!25}{white}{
    \begin{minipage}{.91\textwidth}
    \begin{algorithmic}[1]
        \INPUT FSM $M=\langle Q, \Sigma, q_0, F, \delta\rangle$
        \PRINT{The Church Booleans, Church Boolean functions, and Mairson's fanout gate, as written in lines 1--7 of \cref{listing:church:a4}}
        \STATE{Construct FSM $M'=\langle Q', \Sigma, q_0, F, \delta'\rangle$ from $M$ by making the transition function $\delta$ total and adding states until $|Q'|$ is a power of two (both using sink states)}
        \LET{$n$}{$\log(|Q'|)$}
        \LET{$v$}{$\langle b_n, \ldots, b_2, b_1 \rangle$}
        \LET{$\beta$}{a mapping $Q'\rightarrow \{0, 1\}^n$ that assigns a unique vector for each state}
        \PRINT{\inline[style=sml,literate=*{C}{\textasciicircum}{1}]{fun CC f' = f'\ $\beta(q_0)$}}
        \FORALL{$\sigma\in\Sigma$}
            \LET{$\boldsymbol{f}^\sigma$}{a mapping $\{0, 1\}^n\rightarrow\{0, 1\}^n$ such that $\boldsymbol{f}^\sigma(\beta(q))=\beta(p) \Leftrightarrow \delta_\sigma(q)=p$}
            \PRINT{\inline[style=sml]{fun\ $\sigma$\ $v$ f' = f'\ $\boldsymbol{f}^\sigma(v)$}}
        \ENDFOR
        \LET{$g^F$}{the (one and only) Boolean function such that $g^F(\beta(q))=1 \Leftrightarrow q \in F$}
        \PRINT{\inline[style=sml,literate=*{D}{\textdollar}{1}]{fun D\ $v$ = ($(g^F(v))$ 0 "") + 0}}
    \end{algorithmic}
    \end{minipage}}
\end{algorithm}

As before, the code excerpts in \cref{algorithm:church} contain mathematical symbols.
These are rendered in text as follows:
\begin{itemize}
    \item A Boolean bit $b_i$ is rendered as \inline{T} if $b_i=1$, as \inline{F} if $b_i=0$, or as \inline{bi} if it is a variable (lines 9 and 11).
    \item A Boolean vector $v=b_n\cdots b_2b_1$ is rendered as a tuple \inline[literate=*{E}{\hbox to 1em{.\hss.\hss.}}{1}]{($b_n$, E,\ $b_2$,\ $b_1$)}.
    \item A Boolean vector function $g(v)$ is rendered as a circuit of Church Booleans using Mairson's fanout gate, as described above.
    \item A Boolean vector function $\boldsymbol{g}(v)$ decomposed to $\langle g_n(v), \ldots, g_2(v), g_1(v)\rangle$ is rendered as a tuple \inline[literate=*{E}{\hbox to 1em{.\hss.\hss.}}{1}]{($g_n(v)$, E,\ $g_2(v)$,\ $g_1(v)$)}.
\end{itemize}

\section{Tabulation}
\label{section:tabulation}
The last encoding technique, \emph{tabulation}, organizes the FSM transition function in a table (2D array) and accesses it at compile time.
Tabulation is a flexible method that can be extended to enforce non-regular API protocols, as discussed in \cref{section:beyond}.
We show how to use tabulation to encode a fluent API for $(a^4)^*$, and then, in \cref{algorithm:tabulation}, we generalize the construction for any FSM.

Let type \inline[literate=*{D}{\textdollar}{1}]{DD} denote FSM acceptance of the input word, and \inline{()} (unit) denote its rejection: \[
    \inline[style=sml,literate=*{D}{\textdollar}{1}]{datatype DD = DD}
\]
Consider the following infinite expression:
\begin{equation}\label{eq:e}
    e=\text{\inline[style=sml,literate=*{D}{\textdollar}{1}]{(DD, ((), ((), ((), (DD, ((), $\ \ldots$))))))}}
\end{equation}
Expression $e$ is an infinite cons-list of \inline[literate=*{D}{\textdollar}{1}]{DD} and \inline{()}, where \inline[literate=*{D}{\textdollar}{1}]{DD} repeats every four nodes.
We can use $e$ to implement a (right to left) fluent API for $(a^4)^*$:
\begin{JAVA}[style=sml]
fun ^^ = $e$
fun a (_, e') = e'
fun \$ (\$\$, _) = ()
\end{JAVA}
Each call to \inline{a} removes an item from $e$, and the terminal function \inline[literate=*{D}{\textdollar}{1}]{D} can be called only on item \inline[literate=*{D}{\textdollar}{1}]{DD}, i.e., once every four calls to \inline{a}.
In SML, however, infinite expressions with infinite types cannot be described.
For example, the following attempt to define $e$ fails due to circularity: \[
    \inline[style=sml,literate=*{D}{\textdollar}{1}]{val e = let fun e' () = (DD, ((), ((), ((), e' ())))) in e' () end}
\]

We can describe expression $e$ \cref{eq:e} using four expressions as follows:
\begin{equation}\label{eq:es}
    \begin{aligned}
        e_0 &= \text{\inline[literate=*{D}{\textdollar}{1}]{(DD,\ $e_1$)}} \\
        e_1 &= \text{\inline{((),\ $e_2$)}} \\
        e_2 &= \text{\inline{((),\ $e_3$)}} \\
        e_3 &= \text{\inline{((),\ $e_0$)}} \\
        e &= e_0
    \end{aligned}
\end{equation}
We view expressions $e_0$ through $e_3$ as the rows of a $4\times2$ table, where the tuples' left-hand is the first column, and the right-hand side is the second column.
This \emph{tabular} layout of $e$ reveals its tight connection to the underlying $(a^4)^*$ FSM (\cref{figure:fsm:a4}):
A table row $e_i$ encodes the corresponding FSM state $q_i$, where the first column encodes membership in $F$ using \inline{()} ($e_i\not\in F$) and \inline[literate=*{D}{\textdollar}{1}]{DD} ($e_i\in F$), and the second column encodes the output of the transition function $\delta_a(e_i)$.
For instance, expression $e_2=\text{\inline{((),\ $e_3$)}}$ encodes state $q_2$, since $q_2 \not\in F$ and $\delta_a(q_2)=q_3$ (encoded by $e_3$).
Let's explicitly place expressions $e_0$ through $e_3$ in a table:
\begin{equation}\label{eq:es:table}
    \boldsymbol{e}=\langle e_0, e_1, e_2, e_3\rangle
\end{equation}

To untangle the recursive definition of expressions $e_0$ through $e_3$ \cref{eq:es}, we replace each use of expression $e_j$ with its index in table $\boldsymbol{e}$ \cref{eq:es:table}, which is $j$.
Instead of writing index $j$ as an integer, which cannot be evaluated at compile time, we write it as an \emph{indexing function} \inline{I$j$} that returns the $j$\hh entry of a (general) quadruple:
\begin{JAVA}[style=sml]
fun I0 (x0, x1, x2, x3) = x0
fun I1 (x0, x1, x2, x3) = x1
fun I2 (x0, x1, x2, x3) = x2
fun I3 (x0, x1, x2, x3) = x3
val e0 = (\$\$, I1)
val e1 = ((), I2)
val e2 = ((), I3)
val e3 = ((), I0)
val e = (e0, e1, e2, e3)
\end{JAVA}
Now we can encode the API function \inline{a} as follows:
\begin{JAVA}[style=sml]
fun a (_, I) = I e
\end{JAVA}
Applying \inline{a} to expression $e_i=\text{\inline[literate=*{D}{\textdollar}{1}]{($\cdot$, I$j$)}}$ evaluates to the $j$\hh element of quadruple \inline{e}, \inline{I$j$ e}.
Since $j=(i+1)\bmod4$ for every expression $e_i$, this element is $e_{(i+1)\bmod4}$.
Therefore, function \inline{a} correctly encodes the FSM transition function: \[
    \inline{a e$i$} = \inline{e$j$} \iff \delta_a(q_i)=q_j
\]

The full fluent API encoding is shown in \cref{listing:projection:a4}.
The API is followed by several chain examples.

\begin{JAVA}[float=ht,style=sml,caption={A tabulation SML fluent API for the FSM of \cref{figure:fsm:a4}},label={listing:projection:a4}]
datatype \$\$ = \$\$
fun I0 (x0, x1, x2, x3) = x0
fun I1 (x0, x1, x2, x3) = x1
fun I2 (x0, x1, x2, x3) = x2
fun I3 (x0, x1, x2, x3) = x3
val e0 = (\$\$, I1) (* $\color{comment}q_0 \in F, \delta_a(q_0)=q_1$ *)
val e1 = ((), I2) (* $\color{comment}q_1 \not\in F, \delta_a(q_1)=q_2$ *)
val e2 = ((), I3) (* $\color{comment}q_2 \not\in F, \delta_a(q_1)=q_3$ *)
val e3 = ((), I0) (* $\color{comment}q_3 \not\in F, \delta_a(q_1)=q_0$ *)
val e = (e0, e1, e2, e3)
fun ^^ f' = f' e0 (* $\color{comment}q_0$ is the initial state *)
fun a (_, I) f' = f' (I e)
fun \$ (\$\$, _) = ();
(* chain examples: *)
^^ \$; (* OK, $\color{comment}\varepsilon\in L((a^4)^*)$ *)
^^ a a a a \$; (* OK, $\color{comment}a^4\in L((a^4)^*)$ *)
^^ a \$; (* error, $\color{comment}a^1\not\in L((a^4)^*)$ *)
^^ a a \$; (* error, $\color{comment}a^2\not\in L((a^4)^*)$ *)
^^ a a a a a \$; (* error, $\color{comment}a^5\not\in L((a^4)^*)$ *)
\end{JAVA}

A general FSM with $n=|Q|$ states and $m=|\Sigma|$ letters is encoded by an $n \times (m+1)$ table, i.e., each letter is given its own column.
State $q_i$ (a table row) is encoded by the expression
\begin{JAVA}[style=sml]
val ei = ($a$, ($I^{(1)}$, $\ldots$, $I^{(m)}$))
\end{JAVA}
where $a=\text{\inline[literate=*{D}{\textdollar}{1}]{DD}}$ if $q_i \in F$ or \inline{()} otherwise and $I^{(j)}=\text{\inline{Ik}}$ where $\delta(q_i, \sigma_j)=q_k$.
If $\delta(q_i, \sigma_j)$ is not defined, set $I^{(j)}=\text{\inline{()}}$, making compilation fail whenever the underlying FSM gets stuck.
This means that in order to get the early failure property, described in \cref{section:shuffling:improvements}, we simply need to remove from the FSM every state that cannot reach an accepting state.

The general tabulation encoding algorithm is described in \cref{algorithm:tabulation}.

\begin{algorithm}
    \caption{The tabulation algorithm for encoding an FSM as an SML fluent API}
    \label{algorithm:tabulation}
    \centering
    \fcolorbox{black!25}{white}{
    \begin{minipage}{.91\textwidth}
    \begin{algorithmic}[1]
        \INPUT FSM $M=\langle Q, \Sigma, q_0, F, \delta\rangle$
        \PRINT{\inline[style=sml,literate=*{D}{\textdollar}{1}]{datatype DD = DD}}
        \STATE{Enumerate $Q=\{q_1, q_2, \ldots, q_n\}$}
        \FORALL{$j=1, 2, \ldots, n$}
            \PRINT{\inline[style=sml,literate=*{E}{\hbox to 1em{.\hss.\hss.}}{1}]{fun I$j$ (x1, x2, E, x$n$) = x$j$}}
        \ENDFOR
        \STATE{Enumerate $\Sigma=\{\sigma_1, \sigma_2, \ldots, \sigma_m\}$}
        \FORALL{$i=1, 2, \ldots, n$}
            \LET{$a$}{\inline[literate=*{D}{\textdollar}{1}]{DD} if $q_i\in F$ or else \inline{()}}
            \FORALL{$j=1, 2, \ldots, m$}
                \LET{$I^{(j)}$}{$\text{\inline{I$k$}}$ where $\delta(q_i, \sigma_j) = q_k$}
                \LET{$I^{(j)}$}{$\text{\inline{()}}$ where $\delta(q_i, \sigma_j)$ is not defined}
            \ENDFOR
            \PRINT{\inline[style=sml,literate=*{E}{\hbox to 1em{.\hss.\hss.}}{1}]{val e$i$ = ($a$,\ ($I^{(1)}$,\ $I^{(2)}$, E,\ $I^{(m)}$))}}
        \ENDFOR
        \PRINT{\inline[style=sml,literate=*{E}{\hbox to 1em{.\hss.\hss.}}{1}]{val e = (e1, e2, E, e$n$)}}
        \PRINT{\inline[style=sml,literate=*{C}{\textasciicircum}{1}]{fun CC f' = f' e$i$} where $q_i$ is the initial state}
        \FORALL{$j=1, 2, \ldots, m$}
            \PRINT{\inline[style=sml,literate=*{E}{\hbox to 1em{.\hss.\hss.}}{1}]{val\ $\sigma_j$ (_, (_, E, I, E, _)) f' = f' (I e)} where the inner tuple has $m$ entries and argument \inline{I} appears in the $j$\hh entry}
        \ENDFOR
        \PRINT{\inline[style=sml,literate=*{D}{\textdollar}{1}{E}{\hbox to 1em{.\hss.\hss.}}{1}]{fun D (DD, _) = ()}}
    \end{algorithmic}
    \end{minipage}}
\end{algorithm}

\section{Beyond Regular Languages}
\label{section:beyond}
In \cref{section:shuffling,section:church,section:tabulation} we
described three methods for encoding FSMs as fluent APIs.
These methods can be used to embed regular DSLs in functional
programming languages.
This leaves us wondering about DSLs that are not regular.
It is actually not hard to come up with a fluent API that
recognizes a non-regular language.
For example, the fluent API shown in \cref{listing:functional:dyck}
recognizes the Dyck language of balanced \texttt{L}/\texttt{R} brackets,
known to be non-regular.
(This is a functional re-implementation of the object-oriented fluent API
in \cref{listing:fluent:dyck}.)

\begin{JAVA}[float=ht,style=sml,caption={An SML fluent API encoding the Dyck language of balanced \texttt{L}/\texttt{R} brackets},label={listing:functional:dyck}]
datatype Z = Z
     and 'x S = S of 'x
fun L x f' = f' (S x)
fun R (S x) f' = f' x
fun ^^ f' = f' Z
fun \$ Z = ()
val w1 = ^^ L R \$ (* compiles, `()' are balanced *)
val w2 = ^^ L L R L R R \$ (* compiles, `(()())' are balanced *)
val w3 = ^^ L L R L R L \$ (* does not compile, `(()()(' are not balanced *)
\end{JAVA}

The current section introduces two fluent API encoding techniques
for non-regular languages.
The first technique integrates several simple APIs (as seen in \cref{listing:functional:dyck})
into one, more complex API.
The second technique generalizes our tabulation encoding
(presented in \cref{section:tabulation}) to support all DCFLs.

\subsection{Product APIs}

Consider the fluent APIs \inline{A2} and \inline{A3} shown in \cref{listing:original:apis}.
These APIs encode the languages $(a^2)^*$ and $(a^3)^*$, respectively, using shuffle encoding (cf.~\cref{listing:shuffle:a4}).

\begin{JAVA}[float=ht,style=sml,caption={SML APIs for the languages $(a^2)^*$ and $(a^3)^*$},label={listing:original:apis}]
datatype t = T and f = F
structure A2 = struct
	fun ^^ f' = f' (T, F)
	fun a (x1, x2) f' = f' (x2, x1)
	fun \$ (T, _) = ()
end
structure A3 = struct
	fun ^^ f' = f' (T, F, F)
	fun a (x1, x2, x3) f' = f' (x3, x1, x2)
	fun \$ (T, _, _) = ()
end
\end{JAVA}

We create the \emph{product} of APIs \inline{A2} and \inline{A3} by pairing the inputs and outputs
of each API function, as shown in \cref{listing:product:api}.
For instance, the expression \inline{((T, F), (T, F, F))} (line 2) is a tuple of the
respective expressions \inline{(T, F)} in \inline{A2} and \inline{(T, F, F)} in \inline{A3}.
The resulting API \inline{A6} simulates the two original APIs simultaneously:
the simulation of \inline{A2} is managed on the left-hand side of each tuple and \inline{A3} on the right-hand side.

\begin{JAVA}[float=ht,style=sml,caption={The product of the APIs in \cref{listing:original:apis}},label={listing:product:api}]
structure A6 = struct
	fun ^^ f' = f' ((T, F), (T, F, F))
	fun a ((x1, x2), (x1', x2', x3')) f' = f' ((x2, x1), (x3', x1', x2'))
	fun \$ ((T, _), (T, _, _)) = ()
end
\end{JAVA}

What, accordingly, is the language encoded by \inline{A6}?
Intuitively, since the termination function \inline{$\texttt\$$} enforces the acceptance conditions
of both \inline{A2} and \inline{A3} simultaneously, the language encoded by \inline{A6} is the intersection of their languages: \[
	(a^2)^* \cap (a^3)^* = (a^6)^*
\]
In general, we can use this product method to encode a complex DSL $L$ by
\1 describing it as the intersection of two simpler languages, $L=L_1 \cap L_2$,
\2 encoding these languages as fluent APIs \inline{L1} and \inline{L2},
and \3 writing the product API of \inline{L1} and \inline{L2}.

The \texttt{L}/\texttt{R} API presented at the start of this section (\cref{listing:functional:dyck}) recognizes a non-regular language by
simulating a stack machine (DPDA) at the type level.
We can use similar stack-based APIs in combination with our shuffle, Church, and tabulation encoding methods to create powerful product APIs.
For example, the (enhanced) HTML API discussed in \cref{section:aa} is the product of a shuffle API and several stack machines.
The shuffle API enforces the order of the HTML tags, e.g., that \inline{<html>} can be followed by \inline{<body>} but never by \inline{<p>}, but does not check that each tag has a matching closing tag.
Tag matching is verified in a separate stack-based API, that ignores the ordering of tags.
Yet another stack-based API is used to check that all the rows of an HTML table have the same number of columns.
Although each of these APIs alone encodes a nonsensical language, their product API exactly encodes enhanced HTML, a complex, context-sensitive language.

\subsection{Functional Fluent APIs for DCFLs}

\citet{Gil:2019} presented an algorithm for encoding DCFLs as fluent APIs
in object-oriented languages.
Their algorithm relies on a minimal set of common object-oriented type system features, including
parametric polymorphism and overloading.
For example, a method created by Gil and Roth's construction might look like this:
\begin{JAVA}[language=java]
interface A<x> {
	B<C<x>, x> f();
}
\end{JAVA}
We denote the signature of method \inline{f} as follows: \[
	f: A(x) \rightarrow B(C(x), x)
\]
Note that $f$ explicitly accepts an object of the containing class \inline{A<x>} as input.
There are no additional input parameters in the basic encoding.

In general, object-oriented fluent API methods have the following form:
\begin{equation}\label{eq:function}
	\sigma:\gamma(\boldsymbol{x})\rightarrow\tau
\end{equation}
where $\sigma$ is the method name, $\gamma$ is the name of the containing class, $\boldsymbol{x}$ are its type parameters, and $\tau$ is the return type.
Type $\tau$ is a (possibly) non-grounded type comprising variables $\boldsymbol{x}$ and other API types.
The fluent API also includes an initial variable:
\begin{equation}\label{eq:initial}
	\text{\lstinline/\^\^/}:t_0
\end{equation}
(for ground type $t_0$) and termination methods that return \kk{void} (unit):
\begin{equation}\label{eq:termination}
	\text{\texttt{\textdollar}}:\gamma(\boldsymbol{x})\rightarrow\text{\lstinline/()/}
\end{equation}
Observe that API method signatures \cref{eq:function} are essentially tree rewrite rules.
Thus, we can see a fluent chain of method calls as a series of rewrites of the initial
tree (type) $t_0$ of the initial variable \cref{eq:initial}.
The termination methods \cref{eq:termination} simply separate the API types into
accepting and rejecting types, i.e., by checking whether or not \inline{$\text{\texttt\textdollar}$} accepts type $\gamma$ as input.

For example, the fluent API shown in \cref{listing:fluent:dyck} in \cref{section:preliminaries}, which encodes an \texttt{L}/\texttt{R} Dyck language, defines the following signatures:
\begin{equation}\label{eq:functional:dyck}
	\begin{array}{ll}
		\text{\lstinline/\^\^/}: z & \text{(\textit{a})} \\
		L: z \rightarrow s(z) & \text{(\textit{b})} \\
		L: s(x) \rightarrow s(s(x)) \hspace*{3ex} & \text{(\textit{c})} \\
		R: s(x) \rightarrow x & \text{(\textit{d})} \\
		\text{\texttt{\$}}: z \rightarrow \text{\texttt{()}} & \text{(\textit{e})}
	\end{array}
\end{equation}

The fluent API in \cref{eq:functional:dyck} overloads method $L$ and, therefore, cannot be directly encoded in a functional programming language.
It is definitely possible, however, to write an equivalent functional API---we saw one in \cref{listing:functional:dyck}.
We now describe the construction of another functional fluent API for the \texttt{L}/\texttt{R} Dyck language.
This time, we take a more methodological approach by converting the object-oriented fluent API in \cref{eq:functional:dyck} into a functional fluent API, and then generalize this method for all DCFLs.

The idea is to apply tabulation encoding, presented in \cref{section:tabulation},
on top of Gil and Roth's construction.
Recall that in the FSM case, we encoded each machine state as a table row (tuple) \[
	q\Rightarrow\text{\inline{($a$, ($I^{(1)}$, $\ldots$, $I^{(m)}$))}}
\]
where $a \in \{\text{\inline{()},\inline{$\text{\texttt{\textdollar\textdollar}}$}}\}$ indicates whether state $q$ is accepting and the $I$'s are indexing functions encoding the outgoing $q$ edges.
In the non-regular case, we encode generic types instead of states.
Instead of outgoing edges, we index the rewrite rules of API methods:
\eref[b]{eq:functional:dyck} is $\rho_1$, \eref[c]{eq:functional:dyck} is $\rho_2$,
and \eref[d]{eq:functional:dyck} is $\rho_3$.
In addition, the encoding of generic type $\gamma(\boldsymbol{t})$ recursively includes the encodings of its type arguments $\boldsymbol{t}$.
Thus, type $\gamma(t_1, \ldots, t_k)$ is encoded as follows: \[
	\gamma(t_1, \ldots, t_k)\Rightarrow\text{\inline{($a$, ($I^{(1)}$, $\ldots$, $I^{(m)}$), ($e_1$, $\ldots$, $e_k$))}}
\]
where \1 $a$ indicates whether $\gamma$ is accepting, \2 $I_j$ is the index of $\sigma_j:\gamma(\boldsymbol{x})\rightarrow\tau$ in the table,
and \3 $e_1$ through $e_k$ are the encodings of types $t_1$ through $t_k$.

In our example, types $z$ and $s(x)$ are encoded as follows:
\begin{JAVA}[style=sml]
val z = (\$\$, (I1, ()), ())
fun s x = ((), (I2, I3), (x))
\end{JAVA}
Type $z$ is accepting, $a=\text{\inline{$\text{\texttt{\textdollar\textdollar}}$}}$, by \eref[e]{eq:functional:dyck}, but $s$ is rejecting since function \inline{$\text{\texttt{\textdollar}}$}
does not accept $s$.
Functions \inline{I1}, \inline{I2}, and \inline{I3} are the indices of rules $\rho_1$, $\rho_2$, and
$\rho_3$ of the API methods $L$ and $R$.
Since $z:R\rightarrow\tau$ is not defined, we put \inline{()} in $R$'s locations in the encoding of $z$.
Finally, type $z$ is non-generic and thus its encoding is a value.
On the other hand, $s$ has a type parameter $x$, so we encode it as a function with a single
parameter that captures $x$'s encoding.
Observe that by encoding generic types as functions, we can create the encodings of complex
types by writing them directly as expressions, e.g., the encoding of type $s(s(z))$
is given by the expression \inline{s(s(z))}.

Next, we encode the rewrite rules $\rho_1$, $\rho_2$, and $\rho_3$ as functions and collect them in table \inline{r}:
\begin{JAVA}[style=sml]
fun r1 () = s(z) (* $\text{\eref[b]{eq:functional:dyck}}$ *)
fun r2 (x) = s(s(x)) (* $\text{\eref[c]{eq:functional:dyck}}$ *)
fun r3 (x) = x (* $\text{\eref[d]{eq:functional:dyck}}$ *)
val r = (r1, r2, r3)
\end{JAVA}
These functions do not specify the containing class $\gamma$, but are otherwise identical to the signatures from which they originated \eref[b--d]{eq:functional:dyck}.

Finally, we can encode the fluent API functions.
The initial function simply returns the encoding of the initial type $t_0=z$ \eref[a]{eq:functional:dyck}:
\begin{JAVA}[style=sml]
fun ^^ f' = f' z
\end{JAVA}
Functions $L$ and $R$ are encoded as follows:
\begin{JAVA}[style=sml]
fun L (_, (I, _), X) f' = f' (I r X)
fun R (_, (_, I), X) f' = f' (I r X)
\end{JAVA}
Each function extracts its respective index \inline{I} from the type encoding, applies it to table \inline{r} to get the appropriate rewrite function, and then applies this function to the type arguments \inline{X}.
For instance, when we apply function \inline{L} to encoding \inline{s(z)}, the index is $\inline{I}=\inline{I1}$, the rewrite function is $\inline{I r}=\inline{r2}$,
and the resulting expression $\inline{r2 s(z)}=\inline{s(s(z))}$ correctly encodes the application
of $L$ to type $s(z)$ in the original API \eref[c]{eq:functional:dyck}.
The termination function \inline{$\text{\texttt\textdollar}$} simply verifies that $a=\text{\inline{$\text{\texttt{\textdollar\textdollar}}$}}$:
\begin{JAVA}[style=sml]
fun \$ (\$\$, _, _) = ()
\end{JAVA}

The complete functional fluent API is shown in \cref{listing:functional:dyck:2}.

\begin{JAVA}[float=ht,style=sml,caption={The object-oriented fluent API in \cref{listing:fluent:dyck} and \cref{eq:functional:dyck} converted to SML},label={listing:functional:dyck:2}]
datatype \$\$ = \$\$
fun I1 (x1, x2, x3) = x1
fun I2 (x1, x2, x3) = x2
fun I3 (x1, x2, x3) = x3
val z = (\$\$, (I1, ()), ())
fun s x = ((), (I2, I3), (x))
fun r1 () = s(z)
fun r2 (x) = s(s(x))
fun r3 (x) = x
val r = (r1, r2, r3)
fun ^^ f' = f' z
fun L (_, (I, _), X) f' = f' (I r X)
fun R (_, (_, I), X) f' = f' (I r X)
fun \$ (\$\$, _, _) = ()
\end{JAVA}

The general DCFL encoding algorithm is presented in \cref{algorithm:tabulation:dcfl}.
This algorithm starts by applying Gil and Roth's construction to produce an object-oriented fluent API, and then applies the (modified) tabulation encoding to remove overloading.

\begin{algorithm}
	\caption{The tabulation algorithm for encoding a DPDA as an SML fluent API}
	\label{algorithm:tabulation:dcfl}
	\centering
	\fcolorbox{black!25}{white}{
	\begin{minipage}{.91\textwidth}
    \begin{algorithmic}[1]
        \INPUT DPDA $A$
        \STATE Apply the tree encoding algorithm of \citet{Gil:2019} to convert $A$ into an object-oriented fluent API $P$ as specified in \cref{eq:function,eq:initial,eq:termination}.
        \PRINT{\inline[style=sml,literate=*{D}{\textdollar}{1}]{datatype DD = DD}}
        \STATE{Enumerate $P$'s API functions \cref{eq:function} $R=\{\rho_1, \rho_2, \ldots, \rho_n\}$}
        \FORALL{$j=1, 2, \ldots, n$}
            \PRINT{\inline[style=sml,literate=*{E}{\hbox to 1em{.\hss.\hss.}}{1}]{fun I$j$ (x1, x2, E, x$n$) = x$j$}}
        \ENDFOR
        \STATE{Enumerate $P's$ API methods $\Sigma=\{\sigma_1, \sigma_2, \ldots, \sigma_m\}$}
        \STATE{Enumerate $P$'s API types $\Gamma=\{\gamma_1, \gamma_2, \ldots, \gamma_l\}$}
        \FORALL{$i=1, 2, \ldots, l$}
            \LET{$a$}{\inline[literate=*{D}{\textdollar}{1}]{DD} if $P$ includes the method $\text{\texttt{\textdollar}}:\gamma_i(\boldsymbol{x})\rightarrow\text{\texttt{()}}$ or else \inline{()}}
            \FORALL{$j=1, 2, \ldots, m$}
                \LET{$I^{(j)}$}{$\text{\inline{I$k$}}$ where $\rho_k=(\sigma_j:\gamma_i(\boldsymbol{x})\rightarrow \tau)\in P$}
                \LET{$I^{(j)}$}{$\text{\inline{()}}$ where $(\sigma_j:\gamma_i(\boldsymbol{x})\rightarrow \tau)$ is not defined in $P$}
            \ENDFOR
            \IF{$\gamma_i$ is not generic}
                \PRINT{\inline[style=sml,literate=*{E}{\hbox to 1em{.\hss.\hss.}}{1}]{val\ $\gamma_i$ = ($a$,\ ($I^{(1)}$,\ $I^{(2)}$, E,\ $I^{(m)}$), ())}}
            \ELSE
                \PRINT{\inline[style=sml,literate=*{E}{\hbox to 1em{.\hss.\hss.}}{1}]{fun\ $\gamma_i$ ($\boldsymbol{x}$) = ($a$,\ ($I^{(1)}$,\ $I^{(2)}$, E,\ $I^{(m)}$), ($\boldsymbol{x}$))}}
            \ENDIF
        \ENDFOR
        \FORALL{$i=1, 2, \ldots, n$}
            \LET{$(\sigma:\gamma(\boldsymbol{x})\rightarrow\tau)$}{$\rho_i$}
            \PRINT{\inline{r$i$ ($\boldsymbol{x}$) =\ $\tau$}}
        \ENDFOR
        \PRINT{\inline[style=sml,literate=*{E}{\hbox to 1em{.\hss.\hss.}}{1}]{val r = (r1, r2, E, r$n$)}}
        \PRINT{\inline[style=sml,literate=*{C}{\textasciicircum}{1}]{fun CC f' = f'\ $t_0$} where $\text{\lstinline/\^\^/}:t_0$ is $P$'s initial variable}
        \FORALL{$j=1, 2, \ldots, m$}
            \PRINT{\inline[style=sml,literate=*{E}{\hbox to 1em{.\hss.\hss.}}{1}]{val\ $\sigma_j$ (_, (_, E, I, E, _), X) f' = f' (I r X)} where the inner tuple has $m$ entries and argument \inline{I} appears in the $j$\hh entry}
        \ENDFOR
        \PRINT{\inline[style=sml,literate=*{D}{\textdollar}{1}{E}{\hbox to 1em{.\hss.\hss.}}{1}]{fun D (DD, _, _) = ()}}
    \end{algorithmic}
    \end{minipage}}
\end{algorithm}

\Cref{algorithm:tabulation:dcfl} accepts as input a DPDA accepting the target DCFL.
In line 1, it converts the input into an object-oriented fluent API using tree encoding.
In lines 8 through 16 the algorithm encodes the API types as functions and variables.
The tree rewrite rules are encoded in lines 17 through 20.
The tree pattern $\tau$ is printed as-is.
Finally, the algorithm prints the API functions in lines 21 through 24.

\section{Discussion}
\label{section:zz}
\subsection{Functional vs.~Object-Oriented Fluent APIs}

Functional fluent APIs share many qualities with their object-oriented counterparts.
They provide a fluent DSL embedding method that employs a simple and uniform syntax devoid of host language control structures.
The simplicity of fluent API designs makes them compatible with a wide range of functional host languages.
Our encoding algorithms in \cref{section:shuffling,section:church,section:tabulation,section:beyond} show that functional fluent API constructions can be generalized to entire classes of DSLs, namely all regular languages and DCFLs.
These results are in line with recent advances in the study of object-oriented fluent APIs, focusing on fluent APIs for DCFLs \cite{Gil:2016,Gil:2019,Yamazaki:2019}.

Nevertheless, there are two notable differences between object-oriented and functional fluent APIs.
First, the functional fluent style is cleaner because functions are called without the dot and parenthesis (\inline{x.f()}) common to object-oriented languages.
This removes the last bit of host language syntax from the EDSL.
In combination with operator overloading, found in many popular functional languages, functional fluent APIs can produce extremely accurate DSL embeddings, as demonstrated by our HTML example in \cref{section:aa}.

The second difference relates to embedding of DSL literals.
DSL literals are embedded as function arguments in the fluent API.
In our embedded HTML document in \cref{listing:html}, for example, attribute \inline{src} in the \inline{img} tag receives a string argument specifying the image source:
\begin{JAVA}[numbers=none]
<img src "https://tinyurl.com/yosemite5"/>
\end{JAVA}
Although adding parameters to our fluent API designs is a technical matter, it has an inherent limitation.
The lack of overloading in functional settings means that the fluent APIs cannot overload argument types.
For instance, \inline{img} can receive a string \emph{or} a URL object, but not both.
This detail could possibly hinder certain EDSL designs and should be kept in mind when making functional fluent APIs.

\subsection{A Plethora of Functional Fluent API Encoding Methods}

\Cref{section:shuffling,section:church,section:tabulation} present no less than three different algorithms for encoding regular languages as functional fluent APIs.
The most promising method is probably bit shuffling, described in \cref{section:shuffling}.
Bit shuffling fluent APIs do not employ unification, so they are expected to induce reasonable compilation times across various languages and compilers.
The bit shuffling improvements described in \cref{section:shuffling:improvements} greatly enhance utility by reducing code size and improving error reporting.

The crux of the bit shuffling construction is to describe the FSM as a collection of circuits or logic gates.
In \cref{section:church} we showed how to implement these circuits at the type level using Church Booleans and \citets{Mairson:2004} fanout gate.
Compile-time circuits may have some additional implications for API design beyond fluent API.
For instance, they can be used to conduct numeric calculations in a type-safe physical dimensions library \cite{Kennedy:1994}.

Tabulation encoding (\cref{section:tabulation}) is a flexible embedding method that can be generalized for non-regular DSLs (\cref{section:beyond}).
It provides the theoretical basis for a functional fluent API generator in the style of Fling \cite{Gil:2019,Roth:2019} and TypeLevelLR \cite{Yamazaki:2019}.
Such tools compile LL, LALR, and LR grammars into fluent APIs that validate the DSL syntax at compile time and produce abstract syntax trees (ASTs) at run time.

\subsection{Functional Subchaining}

A fluent API embeds a DSL program as a single chain of method calls.
This design can be limiting when we want to create \emph{excerpts} of the DSL program.
In the HTML document of \cref{listing:html}, we might decide, for example, to place the HTML table in a variable and then reference it in the complete webpage.
Recently, \citet{Yamazaki:2022} presented a new generation method for object-oriented fluent APIs that supports \emph{subchains}---fluent chains that embed DSL fragments.
Since our functional fluent API designs use curried functions, they come with subchaining (almost) out of the box.
\Cref{listing:subchaining} demonstrates functional subchaining using our HTML fluent API.

\begin{JAVA}[float=ht,style=sml,caption={Subchaining with a functional fluent API},label={listing:subchaining}]
fun sc a b = b a (* subchaining helper function *)
val table = fn f' => sc f' (* HTML table subchain *)
	<table>
		<tr> <th> `"Image" </th> </tr>
		<tr> <td> <img src "https://tinyurl.com/yosemite5"/> </td> </tr>
	</table>
val webpage = ^^ (* using the subchain in a complete embedding *)
    <html>
    	<body>
    		table (* subchain usage *)
    	</body>
    </html>
\$\$
\end{JAVA}

Function \inline{sc}, defined in line 1 of \cref{listing:subchaining}, is an auxiliary function for creating subchains.
Our fluent API designs require no additions or modifications to support subchaining besides \inline{sc}.
In lines 2 through 6, we define an HTML excerpt containing an HTML table.
The boilerplate code
\begin{quote}
    \centering
    \inline[style=sml]{fn f' => sc f'}
\end{quote}
in line 2 must precede every subchain.
Variable \inline{table} containing the HTML table is used as part of a complete HTML embedding in lines 7 through 13.

At the type level, using a subchain is equivalent to inlining its code, so the fluent API correctly reports API misuses in subchains.
However, the error messages are displayed where the subchain is referenced and not at the error's actual location.
We speculate that more sophisticated fluent API designs could force the compiler to check the correctness of subchains at the place of definition.

\subsection{Implementation and Experiments}

This work is accompanied by \emph{Flunct} \cite{Roth:2023:artifact}, an SML implementation of the bit shuffling, Church, and tabulation encoding techniques.
Flunct compiles a given FSM $M$ into a functional fluent API that validates $M$ at compile time.
\ifdefined\FullVersion
In \cref{section:flunct} we demonstrate how Flunct can be used to embed a DSL in SML.
We used Flunct to conduct a series of experiments to measure the code size and compilation time of the various fluent API constructions described in the paper.
The results of these experiments are described in \cref{section:experiments}.
\else
In the full paper \cite{Roth:2023}, we demonstrate how Flunct can be used to embed a DSL in SML.

We used Flunct to conduct a series of experiments to measure the code size and compilation time of the various fluent API constructions described in the article.
For example, the graph in \cref{figure:ring:time:1} compares the compilation times of bit shuffling, Church, and tabulation fluent APIs using the MLton SML compiler.
Our experiments suggest that bit shuffling, together with the improvements described in \cref{section:shuffling:improvements}, performs the best in terms of code size and compilation time, which is expected considering that bit shuffling is the only technique that doesn't rely on unification.
A full description of the experiments is found in the full paper.

\begin{figure}[ht]
  \centering
  \begin{tikzpicture}
	\begin{loglogaxis}[
	  xlabel={Chain size (\#calls)},
	  ylabel={Compilation time (seconds)},
	  legend pos=outer north east,
	  log basis x=2,
	  log basis y=2,
	]
	\addplot[
	  color=green!50!black,
	  mark=x,
	] coordinates {
		(0.5,2.897)(16,2.837)(32,2.838)(64,2.853)(128,3.024)(256,2.584)(512,2.267)(1024,2.358)(2048,2.830)(4096,2.771)(8192,4.051)(16384,5.695)(32768,13.979)
	};
	\addlegendentry{\shortstack[l]{Bit Shuffling (\cref{section:shuffling}) \\[-2pt] + improvements (\cref{section:shuffling:improvements})}}
	\addplot[
	  color=red,
	  mark=square,
	] coordinates {
		(0.5,2.500)(16,3.205)(32,3.151)(64,3.566)(128,5.095)(256,8.427)(512,15.324)
	};
	\addlegendentry{Church (\cref{section:church})}
	\addplot[
	  color=blue,
	  mark=triangle,
	] coordinates {
		(0.5,3.068)(16,3.092)(32,3.029)(64,3.016)(128,3.306)(256,3.064)(512,3.545)(1024,4.661)(2048,5.256)(4096,8.910)(8192,15.354)
	};
	\addlegendentry{Tabulation (\cref{section:tabulation})}
	\end{loglogaxis}
  \end{tikzpicture}
  \caption{Chain length vs.~compilation time of bit shuffling, Church, and tabulation fluent APIs using MLton}
  \label{figure:ring:time:1}
\end{figure}
\fi

\ifdefined\FullVersion\newpage\fi

\bibliography{00.bib}

\ifdefined\FullVersion

\newpage

\appendix

\section{Flunct---a Functional Fluent API Generator}
\label{section:flunct}
Flunct embeds a DSL specified by an FSM as a functional fluent API using the bit shuffling, Church Boolean, and tabulation techniques described in \cref{section:shuffling,section:church,section:tabulation}.
In this appendix, we show how to use Flunct to embed a DSL in SML.
The running example DSL is the \emph{Command-Line Interface (CLI) Builder}, a DSL for creating command-line applications.
For instance, we can use the CLI Builder DSL to create an ``ice cream machine'' command-line application:
\begin{JAVA}[caption={The ice cream machine CLI},label={listing:icecream}]
icecream --flavors chocolate vanilla --container cup > ice1.cream
icecream --flavors pistachio --toppings caramel  > ice2.cream
icecream --help
¢\color{black!60}Ice Cream Machine¢
¢\color{black!60}\ \ Makes customized ice cream.¢
¢\color{black!60}Options:¢
¢\color{black!60}\ \ --flavors, -f [list]¢
¢\color{black!60}\ \ \ \ list of ice cream flavors¢
¢$\color{black!60}\ldots$¢
\end{JAVA}

The CLI Builder supports the following properties:
\begin{enumerate}
    \item \texttt{name}: Application name.
    \item \texttt{description}: Application description (optional).
    \item A list of zero or more arguments comprising:
    \begin{enumerate}
        \item \texttt{argument}: Argument name.
        \item \texttt{optional}: Makes the argument optional (optional).
        \item \texttt{alias}: A list of zero or more argument aliases.
        \item \texttt{argtype}: The type of values accepted by the argument (optional); has a list of zero or more type aliases:
        \begin{enumerate}
            \item \texttt{alias}: If specified, the argument type is an enumeration of all type aliases.
        \end{enumerate}
        \item \texttt{description}: Argument description (optional).
    \end{enumerate}
\end{enumerate}
We can formally define the CLI Builder DSL using a regular expression:
\begin{equation*}
    \texttt{name description? (optional? argument alias$^*$ (argtype alias$^*$)? description)$^*$}
\end{equation*}
Flunct does not support regular expressions at present.
Hence, the first step in embedding the CLI Builder DSL is to encode it as an FSM, as shown in \cref{figure:cli:fsm}.

\begin{figure}[ht]
    \centering
    \begin{tikzpicture}[node distance=10ex]
        \node[state,initial] (q0) at (0,0) {$q_0$};
        \node[state,accepting] (q1) at (2.5,0) {$q_1$};
        \node[state,accepting] (q2) at (5,0) {$q_2$};
        \node[state] (q3) at (7.5,0) {$q_3$};
        \node[state,accepting] (q4) at (7.5,-2.5) {$q_4$};
        \node[state,accepting] (q5) at (5,-2.5) {$q_5$};
        \draw[->,opacity=.5,dashed]
            (q1) edge[bend right=20] (q3)
            (q1) edge[bend left=10] (q4)
            (q2) edge[bend left=10] (q4)
            (q4) edge[bend left=10] (q2)
            (q4) edge[bend left=20] (q3)
            (q4) edge[loop below] (q4)
            (q5) edge[bend right=20] (q4)
            (q5) edge (q3)
            ;
        \draw[->]
            (q0) edge node[above]{name} (q1)
            (q1) edge node[above]{description} (q2)
            (q2) edge node[above]{optional} (q3)
            (q3) edge node[right]{argument} (q4)
            (q4) edge[loop right] node[right]{alias} (q4)
            (q4) edge node[above]{argtype} (q5)
            (q5) edge node[left]{description} (q2)
            (q5) edge[loop left] node[left]{alias} (q5);
    \end{tikzpicture}
    \vspace{-1.5ex}
    \caption{An FSM describing the CLI Builder interface. Some edges are grayed-out for readability.}
    \label{figure:cli:fsm}
\end{figure}
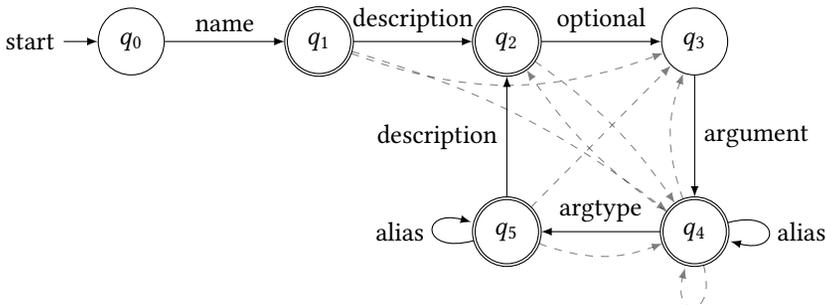

We can use Flunct's interface to describe the FSM of \cref{figure:cli:fsm} in SML.
First, we write the FSM alphabet:
\begin{JAVA}[style=sml]
val name = Letter ("name", ["string"])
val description = Letter ("description", ["string"])
val optional = Letter ("optional", [])
val argument = Letter ("argument", ["string"])
val alias = Letter ("||", ["string"])
val argtype = Letter ("argtype", ["string"])
\end{JAVA}
The \inline{Letter} constructor accepts the API function name and the list of its argument types.
Most functions in our CLI Builder example accept a single \inline{string} argument.
Notice that we chose to encode the \texttt{alias} function as an operator, \inline{||}.

Next, we define the FSM depicted in \cref{figure:cli:fsm}:
\begin{JAVA}[style=sml]
val fsm = Fsm (
	State 0, (* initial state *)
	[State 1, State 2, State 4, State 5], (* accepting states *)
	[ (* FSM transitions *)
		Transition (State 0, name, State 1),
		Transition (State 1, description, State 2),
		Transition (State 1, optional, State 3),
		Transition (State 1, argument, State 4),
		Transition (State 2, optional, State 3),
		Transition (State 2, argument, State 4),
		Transition (State 3, argument, State 4),
		Transition (State 4, description, State 2),
		Transition (State 4, optional, State 3),
		Transition (State 4, alias, State 4),
		Transition (State 4, argument, State 4),
		Transition (State 4, argtype, State 5),
		Transition (State 5, description, State 2),
		Transition (State 5, optional, State 3),
		Transition (State 5, argument, State 4),
		Transition (State 5, alias, State 5)
	]	
)
\end{JAVA}
The FSM includes an initial state (line 2), a set of accepting states (line 3), and a set of FSM transitions (lines 5 to 20).
Each transition comprises a source state, input letter, and a target state.
For example, line 7 defines that the FSM moves from state $q_1$ to state $q_3$ when reading \texttt{optional}.

To convert the FSM into an SML fluent API we simply call
\begin{JAVA}[style=sml,numbers=none]
print (flunct_api "CLI" fsm Sparse)
\end{JAVA}
The last argument specifies the encoding method.
Flunct supports four methods: \inline{Shuffle} (bit shuffling), \inline{Church} (Church Booleans), \inline{Tabulation}, and \inline{Sparse} (improved bit shuffling as per \cref{section:shuffling:improvements}).
Running this line produces a complete fluent API for the CLI builder in \inline[style=sml]{struct CLI}.

We can now define the ice cream machine CLI inside SML:
\begin{JAVA}[style=sml]
val icecream_cli = let
		open CLI
	in
		^^
			name "Ice Cream Machine"
			description "Makes customized ice cream."
			argument "--flavors" || "-f"
				argtype "@list"
				description "list of icecream flavors"
			optional argument "--toppings" || "-t"
				argtype "@list"
				description "list of icecream flavors"
			optional argument "--container" || "-c"
				argtype "cone" || "cup" || "Styrofoam"
		\$\$
	end
\end{JAVA}
The fluent API catches API misuses at compile time.
If, for instance, we write \inline{optional} twice in line 10, the code does not compile:
\begin{JAVA}[language={}]
Error: operator and operand don't agree [tycon mismatch]
...
in expression: ((((<exp> <exp>) description) "list of icecream flavors") optional) optional
\end{JAVA}

The embedded program produces a list of tokens:
\begin{JAVA}[style=sml]
val icecream_cli = [Name "Ice Cream Machine", Description "Makes customized ice cream.",
                    Argument "--flavors", @|| "-f", ...] : call list
\end{JAVA}
When the CLI Builder application is completed, its backend would parse and process these tokens.

\section{Code Size and Compilation Time Experiments}
\label{section:experiments}
This appendix describes a series of experiments measuring the code size and compilation time of the different fluent API constructions described throughout the paper.
Most of the experiments were conducted using Flunct, our functional fluent API generator introduced in \cref{section:flunct}.
The purpose of these experiments is to give us a general idea about the feasibility of fluent APIs in different situations.
In particular, the results may vary when switching the underlying DSLs.

\emph{A logarithmic scale is applied to all graph axes in this appendix.}
All compilation times are averaged over \emph{five} runs.

\subsection{The Ring Experiment}

A \emph{ring} is an FSM with $n$ states that recognizes the language $(a^n)^*$, as depicted in \cref{figure:ring}.
The ring's states are organized in a circle with each state having a single edge to the state on its right, and the initial state being the only accepting state.
For example, the article's running example FSM from \cref{figure:fsm:a4} is a 4-state ring.
We use rings as a simple and scalable FSM design.

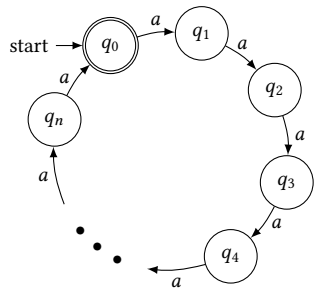
\begin{figure}[h]
	\centering
	\begin{tikzpicture}[scale=.8,every node/.style={scale=.8}]
		\node[state,initial,accepting] (q0) at (120:2) {$q_0$};
		\node[state] (q1) at (75:2) {$q_1$};
		\node[state] (q2) at (30:2) {$q_2$};
		\node[state] (q3) at (-15:2) {$q_3$};
		\node[state] (q4) at (-60:2) {$q_4$};
		\node (d1) at (-105:2) {};
		\node (d2) at (-116.25:2) {$\bullet$};
		\node (d3) at (-127.5:2) {$\bullet$};
		\node (d4) at (-138.75:2) {$\bullet$};
		\node (d5) at (-150:2) {};
		\node[state] (qn) at (-195:2) {$q_n$};
		\draw[->]
			(q0) edge[bend left=10] node[above]{$a$} (q1)
			(q1) edge[bend left=10] node[above]{$a$} (q2)
			(q2) edge[bend left=10] node[right]{$a$} (q3)
			(q3) edge[bend left=10] node[right]{$a$} (q4)
			(q4) edge[bend left=10] node[below]{$a$} (d1)
			(d5) edge[bend left=10] node[left]{$a$} (qn)
			(qn) edge[bend left=10] node[left]{$a$} (q0);
	\end{tikzpicture}
	\caption{A ring FSM with $n$ states}
	\label{figure:ring}
\end{figure}

In the ring experiment, we compiled rings of different sizes into fluent APIs using different encoding methods.
Recall that the \emph{Sparse} encoding method is bit shuffling augmented by the improvements described in \cref{section:shuffling:improvements}.
We then measured the compilation times of these fluent APIs on chains of various length with different compilers and languages.
A \emph{chain} of length $k$ is a fluent expression with $k$ calls to \inline{a}:
\begin{equation*}
	\text{\inline[literate=*{D}{\textdollar}{1}{C}{\textasciicircum}{1}{E}{\hbox to 1em{.\hss.\hss.}}{1}]{CC a a E a DD}}
\end{equation*}

\newpage

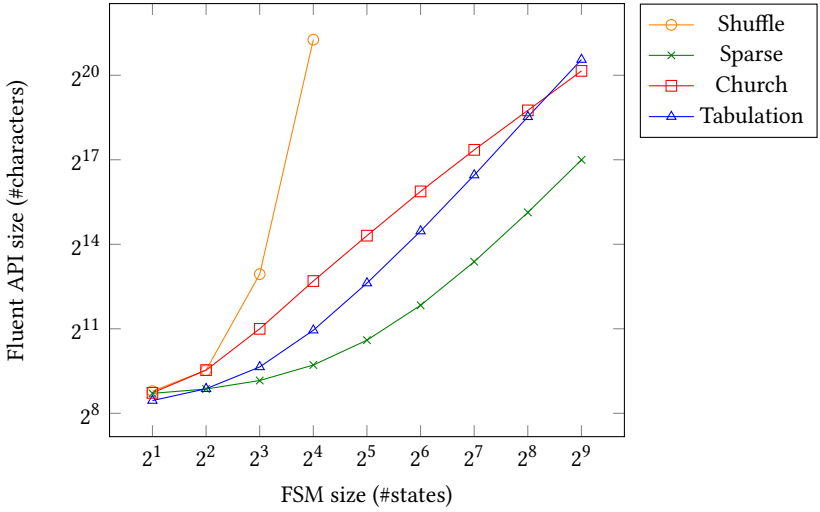
\begin{figure}[h]
  \centering
  \bigskip\bigskip\bigskip\bigskip
  \small
  \begin{tikzpicture}
	\begin{loglogaxis}[
	  xlabel={FSM size (\#states)},
	  ylabel={Fluent API size (\#characters)},
	  legend pos=outer north east,
	  log basis x=2,
	  log basis y=2,
	]
	\addplot[
	  color=orange,
	  mark=o,
	] coordinates {
	  (2,441)(4,740)(8,7841)(16,2511805)
	};
	\addlegendentry{Shuffle}
	\addplot[
	  color=green!50!black,
	  mark=x,
	] coordinates {
	  (2,418)(4,465)(8,573)(16,839)(32,1546)(64,3647)(128,10656)(256,35840)(512,130592)
	};
	\addlegendentry{Sparse}
	\addplot[
	  color=red,
	  mark=square,
	] coordinates {
	  (2,423)(4,740)(8,2045)(16,6622)(32,20231)(64,60088)(128,167021)(256,440834)(512,1164947)
	};
	\addlegendentry{Church}
	\addplot[
	  color=blue,
	  mark=triangle,
	] coordinates {
	  (2,350)(4,468)(8,800)(16,1974)(32,6294)(64,22614)(128,89698)(256,376290)(512,1539298)
	};
	\addlegendentry{Tabulation}
	\end{loglogaxis}
  \end{tikzpicture}
  \caption{Ring FSM size vs.~fluent API size}
  \label{figure:ring:size}
  \bigskip\bigskip\bigskip\bigskip\bigskip
\end{figure}

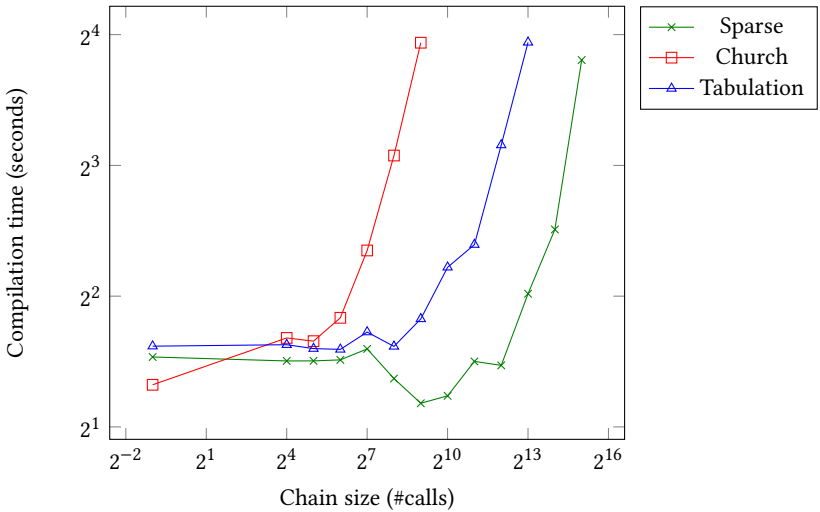
\begin{figure}[h]
  \centering
  \small
  \begin{tikzpicture}
	\begin{loglogaxis}[
	  xlabel={Chain size (\#calls)},
	  ylabel={Compilation time (seconds)},
	  legend pos=outer north east,
	  log basis x=2,
	  log basis y=2,
	]
	\addplot[
	  color=green!50!black,
	  mark=x,
	] coordinates {
		(0.5,2.897)(16,2.837)(32,2.838)(64,2.853)(128,3.024)(256,2.584)(512,2.267)(1024,2.358)(2048,2.830)(4096,2.771)(8192,4.051)(16384,5.695)(32768,13.979)
	};
	\addlegendentry{Sparse}
	\addplot[
	  color=red,
	  mark=square,
	] coordinates {
		(0.5,2.500)(16,3.205)(32,3.151)(64,3.566)(128,5.095)(256,8.427)(512,15.324)
	};
	\addlegendentry{Church}
	\addplot[
	  color=blue,
	  mark=triangle,
	] coordinates {
		(0.5,3.068)(16,3.092)(32,3.029)(64,3.016)(128,3.306)(256,3.064)(512,3.545)(1024,4.661)(2048,5.256)(4096,8.910)(8192,15.354)
	};
	\addlegendentry{Tabulation}
	\end{loglogaxis}
  \end{tikzpicture}
  \caption{Chain length vs.~compilation time with 16-state Ring FSM and MLton}
  \label{figure:ring:time:1}
\end{figure}

\newpage

\begin{figure}[h]
  \centering
  \bigskip\bigskip\bigskip\bigskip
  \small
  \begin{tikzpicture}
	\begin{loglogaxis}[
	  xlabel={FSM size (\#states)},
	  ylabel={Compilation time (seconds)},
	  legend pos=outer north east,
	  log basis x=2,
	  log basis y=2,
	]
	\addplot[
	  color=green!50!black,
	  mark=x,
	] coordinates {
		(2,2.871)(4,2.742)(8,2.471)(16,3.015)(32,2.269)(64,2.628)(128,2.818)(256,2.951)(512,2.764)
	};
	\addlegendentry{Sparse}
	\addplot[
	  color=red,
	  mark=square,
	] coordinates {
		(2,3.644)(4,2.913)(8,2.871)(16,2.348)(32,2.835)(64,6.743)(128,27.116)
	};
	\addlegendentry{Church}
	\addplot[
	  color=blue,
	  mark=triangle,
	] coordinates {
		(2,2.693)(4,2.787)(8,2.998)(16,3.131)(32,3.153)(64,3.117)(128,2.681)(256,5.277)(512,14.298)
	};
	\addlegendentry{Tabulation}
	\end{loglogaxis}
  \end{tikzpicture}
  \caption{Ring FSM size vs.~compilation time of the empty chain (\protect\inline[literate=*{D}{\textdollar}{1}{C}{\textasciicircum}{1}]{CC DD}) with MLton}
  \label{figure:ring:time:2}
  \bigskip\bigskip\bigskip\bigskip\bigskip\bigskip
\end{figure}

\begin{figure}[h]
  \centering
  \small
  \begin{tikzpicture}
	\begin{loglogaxis}[
	  xlabel={Chain size (\#calls)},
	  ylabel={Compilation time (seconds)},
	  legend pos=outer north east,
	  log basis x=2,
	  log basis y=2,
	]
	\addplot[
	  color=green!50!black,
	  mark=x,
	] coordinates {
	  (0.5,0.001)(4,0.001)(16,0.031)(64,2.289)
	};
	\addlegendentry{Sparse}
	\addplot[
	  color=red,
	  mark=square,
	] coordinates {
	  (0.5,0.002)(4,0.681)
	};
	\addlegendentry{Church}
	\addplot[
	  color=blue,
	  mark=triangle,
	] coordinates {
	  (0.5,0.001)(4,0.001)(16,0.008)(64,0.581)
	};
	\addlegendentry{Tabulation}
	\end{loglogaxis}
  \end{tikzpicture}
  \caption{Chain length vs.~compilation time with 4-state Ring FSM and SML/NJ}
  \label{figure:ring:time:1:smlnj}
\end{figure}
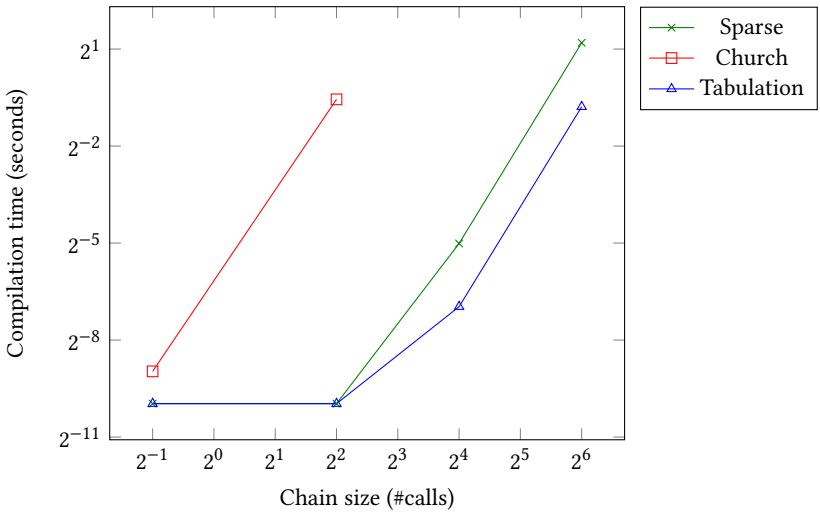

\newpage

\Cref{figure:ring:size} depicts the fluent APIs' code size.
As expected, the size of bit shuffling fluent APIs is exponential in the ring size.
There are two variables to be considered when measuring compilation time: the FSM size and chain length.
\Cref{figure:ring:time:1} shows the compilation times of chains of various lengths using the MLton SML compiler and a constant 16-state ring.
\Cref{figure:ring:time:2} depicts the compilation times of the empty chain (zero length) on rings of various sizes using MLton.
The SML/NJ interpreter is not quite suitable for these kind of experiment since it cannot function as a compiler.
Nevertheless, it is possible to approximate the compilation time of SML/NJ on \emph{individual expressions}.
\Cref{figure:ring:time:1:smlnj} shows the compilation times of chains of various lengths using the SML/NJ interpreter and a constant 4-state ring.
These compilation times are calculated on the chains alone, without the fluent APIs.
Note that SML/NJ gets stuck for quite some time when compiling chains longer than the ones used in \cref{figure:ring:time:1:smlnj}.

\subsection{The HTML Experiment}

In this experiment we measured the compilation time of our HTML API, introduced in \cref{section:aa}.
As explained in \cref{section:beyond}, the HTML fluent API is the product of a bit shuffling API with a 61-state FSM and several stack machines.
We used this API to embed the HTML webpage in \cref{listing:html} and added the following table after line 15:
\begin{JAVA}[style=sml,firstnumber=16]
<table>
    <tr> <th> `"nested table" </th> </tr>
    <tr>
        <td>
            (nested_table (table_depth - 1))
        </td>
    </tr>
</table>
\end{JAVA}
The expression \inline{(inner_table (table_depth - 1))} (line 20) was replaced by an identical nested table of variable depth, yielding increasingly complex embeddings.
A table of zero depth was replaced by
\begin{JAVA}[style=sml,numbers=none]
`"recursion!"
\end{JAVA}
\Cref{figure:html:time} depicts the compilation time of HTML documents with tables of variable nesting levels using MLton.

\subsection{The DCFL Experiment}

In this experiment we measured the compilation time of the fluent API in \cref{listing:functional:dyck:2} encoding a DCFL.
Recall that this fluent API was created using (extended) tabulation encoding as described in \cref{section:beyond}.
We used chains of variable lengths and repeated the experiment in two languages:
Once in SML with MLton and once in the ELM programming language \cite{Czaplicki:2013} using the ELM compiler.
The results are shown in \cref{figure:dcfl:time}.

\subsection{Technical Details}

The experiments described in this appendix were conducted on a (fairly old) laptop with the following properties:
\begin{itemize}
    \item Operating system: Ubuntu 20.04.
    \item Processor: Intel(R) Core(TM) i7-5600U CPU @ 2.60GHz
    \item MLton version: 20130715
    \item SML/NJ version: v110.79
    \item ELM version: 0.19.1
\end{itemize}

\newpage

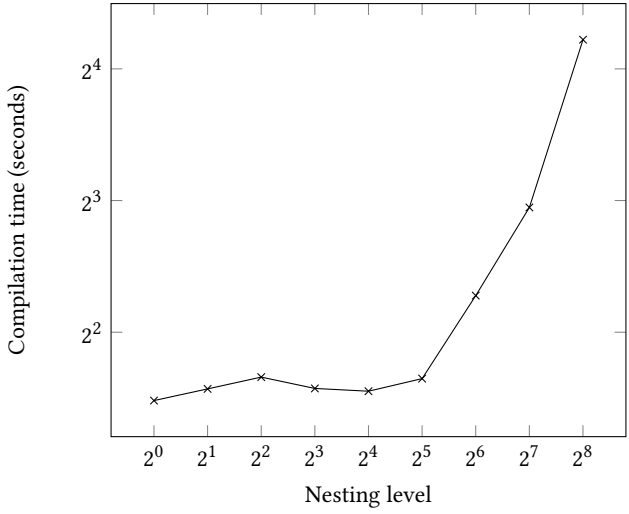
\begin{figure}[h]
  \centering
  \bigskip\bigskip\bigskip\bigskip
  \small
  \begin{tikzpicture}
	\begin{loglogaxis}[
	  xlabel={Nesting level},
	  ylabel={Compilation time (seconds)},
	  legend pos=outer north east,
	  log basis x=2,
	  log basis y=2,
	]
	\addplot[
	  color=black,
	  mark=x,
	] coordinates {
	  (0,2.317)(1,2.791)(2,2.968)(4,3.158)(8,2.975)(16,2.932)(32,3.133)(64,4.850)(128,7.714)(256,18.653)
	};
	\end{loglogaxis}
  \end{tikzpicture}
  \caption{Example HTML size vs.~compilation time with MLton}
  \label{figure:html:time}
  \bigskip\bigskip\bigskip\bigskip\bigskip\bigskip
\end{figure}

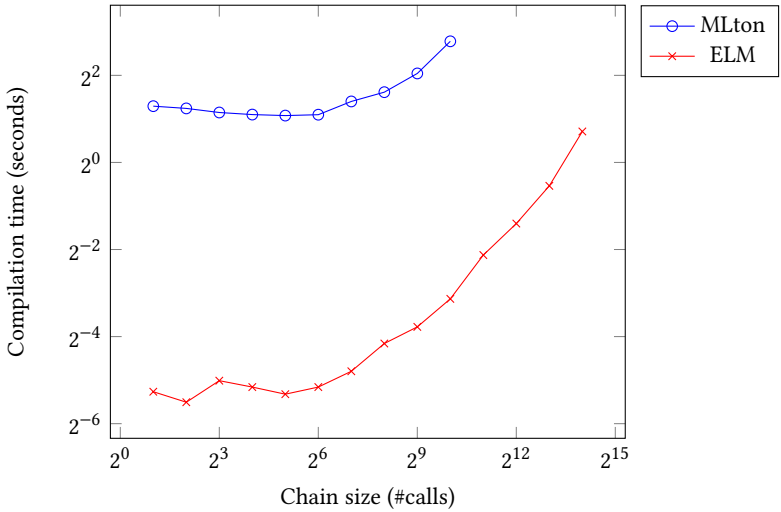
\begin{figure}[h]
  \centering
  \small
  \begin{tikzpicture}
	\begin{loglogaxis}[
	  xlabel={Chain size (\#calls)},
	  ylabel={Compilation time (seconds)},
	  legend pos=outer north east,
	  log basis x=2,
	  log basis y=2,
	]
	\addplot[
	  color=blue,
	  mark=o,
	] coordinates {
	  (2,2.449)(4,2.363)(8,2.214)(16,2.142)(32,2.108)(64,2.138)(128,2.640)(256,3.060)(512,4.123)(1024,6.875)
	};
	\addlegendentry{MLton}
	\addplot[
	  color=red,
	  mark=x,
	] coordinates {
	  (2,0.026)(4,0.022)(8,0.031)(16,0.028)(32,0.025)(64,0.028)(128,0.036)(256,0.056)(512,0.073)(1024,0.114)(2048,0.229)(4096,0.378)(8192,0.689)(16384,1.637)
	};
	\addlegendentry{ELM}
	\end{loglogaxis}
  \end{tikzpicture}
  \caption{Chain length vs.~compilation time when compiling the fluent API in \protect\cref{listing:functional:dyck:2} in SML and ELM}
  \label{figure:dcfl:time}
\end{figure}

\fi

\end{document}